\documentclass{ws-ijmpa}
\usepackage[super,compress]{cite}
\usepackage{graphicx}
\usepackage{braket}
\usepackage{bm}
\usepackage{comment}
\usepackage{here}

\begin{document}
\markboth{Teruyuki Kitabayashi}{Primordial black holes and scotogenic dark matter}

\def\Journal#1#2#3#4{{#1} {\bf #2}, #3 (#4)}
\def\AHEP{Advances in High Energy Physics.} 
\def\ARNPS{Annu. Rev. Nucl. Part. Sci.} 
\def\AandA{Astron. Astrophys.} 
\def\ANP{Ann. Phys.}
\def\APJ{Astrophys. J.}
\def\APJS{Astrophys. J. Suppl}
\def\CMP{Commn. Math. Phys.}
\def\COMR{Comptes Rendues}
\def\CPC{Chin. Phys. C}
\def\CQG{Class. Quant. Grav.}
\def\EPJC{Eur. Phys. J. C}
\def\EPJP{Eur. Phys. J. Plus}
\def\FASS{Front. Astron. Space Sci.}
\def\IJMPA{Int. J. Mod. Phys. A}
\def\IJMPD{Int. J. Mod. Phys. D}
\def\IJMPE{Int. J. Mod. Phys. E}
\def\JCAP{J. Cosmol. Astropart. Phys.}
\def\JHEP{J. High Energy Phys.}
\def\JETPL{JETP. Lett.}
\def\JETPUSSR{JETP (USSR)}
\def\JPG{J. Phys. G} 
\def\JPGNP{J. Phys. G: Nucl. Part. Phys.} 
\def\MPLA{Mod. Phys. Lett. A}
\def\MNRAS{Mon. Not. R. Astron. Soc.}
\def\NIMA{Nucl. Instrum. Meth. A.}
\def\NATU{Nature}
\def\NATULONDON{Nature (London)}
\def\NCA{Nuovo Cimento}
\def\NJP{New. J. Phys.}
\def\NPB{Nucl. Phys. B}
\def\NPBOLD{Nucl. Phys.}
\def\NPBSUPPL{Nucl. Phys. B. Proc. Suppl.}
\def\PLB{{Phys. Lett.} B}
\def\PMCA{PMC Phys. A}
\def\PREP{Phys. Rep.}
\def\PPNP{Prog. Part. Nucl. Phys.}
\def\PLBOLD{Phys. Lett.}
\def\PAN{Phys. Atom. Nucl.}
\def\PRL{Phys. Rev. Lett.}
\def\PRD{Phys. Rev. D}
\def\PRC{Phys. Rev. C}
\def\PR{Phys. Rev.}
\def\PTP{Prog. Theor. Phys.}
\def\PTEP{Prog. Theor. Exp. Phys.}
\def\RMP{Rev. Mod. Phys.}
\def\RPP{Rep. Prog. Phys.}
\def\SJNP{Sov. J. Nucl. Phys.}
\def\SPJETP{Sov. Phys. JETP}
\def\SCIENCE{Science}
\def\TNYAS{Trans. New York Acad. Sci.}
\def\ZETP{Zh. Eksp. Teor. Piz.}
\def\ZFPH{Z. fur Physik}
\def\ZPC{Z. Phys. C}

%
\catchline{}{}{}{}{}
%


\title{Primordial black holes and scotogenic dark matter}

\author{Teruyuki Kitabayashi}

\address{Department of Physics, Tokai University,\\
4-1-1 Kitakaname, Hiratsuka, Kanagawa 259-1292, Japan\\
teruyuki@tokai-u.jp}

\maketitle

\begin{history}
\received{Day Month Year}
\revised{Day Month Year}
\end{history}

\begin{abstract}
We study the effect of the scotogenic dark matter on the primordial black holes (PBHs) and vice versa. We show that if the PBHs evaporate in the radiation dominant era, the upper limit of the initial mass of the PBHs $M_{\rm in}$ should be constrained as $10^4 \lesssim M_{\rm in}/M_{\rm Pl} \lesssim  10^{10}$ for $\mathcal{O}$(1) TeV scotogenic dark matter ($\mathcal{O}$(1) TeV is the most appropriate energy scale in the scotogenic model). On the other hand, if the PBHs evaporate in the PBH dominated era, a quite heavy scotogenic dark matter ($m_{\rm DM} \gtrsim 10^9$ GeV) for $M_{\rm in}/M_{\rm Pl} \sim 10^{13}$ may be allowed. 
\end{abstract}

\ccode{PACS numbers:12.60.-i,14.60.St,95.35.+d,98.80.Cq}


\section{Introduction\label{section:introduction}}
Understanding the nature of both dark matter and neutrinos is one of the big problems in cosmology and particle physics. The so-called scotogenic model is one of the attractive and extensively studied dark matter models \cite{Ma2006PRD}. The scotogenic model can simultaneously account for dark matter candidates and the origin of tiny masses of neutrinos \cite{Ma2006PRD}. In this model, the lightest heavy Majorana particle is a good candidate of the dark matter \cite{Suematsu2009PRD,Suematsu2010PRD}, and neutrino masses are generated by one-loop interactions mediated by the dark matter candidate. The scotogenic model as well as one-loop interactions related to dark matter and neutrino mass have been extensively studied in the literature \cite{Kubo2006PLB,Hambye2007PRD,Farzan2009PRD,Farzan2010MPLA,Farzan2011IJMPA,Kanemura2011PRD,Schmidt2012PRD,Faezan2012PRD,Aoki2012PRD,Hehn2012PLB,Bhupal2012PRD,Bhupal2013PRD,Law2013JHEP,Kanemura2013PLB,Hirsch2013JHEP,Restrepo2013JHEP,Ho2013PRD,Lindner2014PRD,Okada2014PRD89,Okada2014PRD90,Brdar2014PLB,Toma2014JHEP,Ho2014PRD,Faisel2014PRD,Vicente2015JHEP,Borah2015PRD,Wang2015PRD,Fraser2016PRD,Adhikari2016PLB,Ma2016PLB,Arhrib2016JCAP,Okada2016PRD,Ahriche2016PLB,Lu2016JCAP,Cai2016JHEP,Ibarra2016PRD,Lindner2016PRD,Das2017PRD,Singirala2017CPC,Kitabayashi2017IJMPA,AbadaJHEP2018,Baumholzer2018JHEP,Ahriche2018PRD,Hugle2018PRD,Kitabayashi2018PRD,Reig2019PLB,Boer2020PRD,Ahriche2020PRD,Faisel2014PLB}.  

On the other hand, the relation between the primordial black holes (PBHs) \cite{Carr1975APJ} and dark matter is also studied. The PBHs emit particles via the Hawking radiation \cite{Hawking1975CMP}. Since the Hawking radiation is induced by gravity, PBHs evaporate into all particle species. Dark matter is also produced by Hawking radiation of the PBHs. The resultant Hawking radiation of PBHs has been considered as a possible explanation for baryogenesis and/or dark matter production \cite{Bell1999PRD,Green1999PRD,Khlopov2006CQG,Baumann2007arXiv,Dai2009JCAP,Fujita2014PRD,Allahverdi2018PRD,Lennon2018JCAP,Morrison2019JCAP,Hooper2019JHEP,Masina2020EPJP,Baldes2020JCAP,Bernal2020arXiv1,Gondolo2020PRD,Bernal2020arXiv2,Auffinger2020arXiv,Datta2021arXiv,Chaudhuri202011arXiv}. The primordial black holes (PBHs) \cite{Carr1975APJ} are produced in the early Universe via a number of mechanisms. The simplest PBH producing mechanism is the collapse of large density perturbations generated from inflation \cite{Garcia-Bellido1996PRD,Kawasaki1998PRD,Yokoyama1998PRD,Kawasaki2006PRD,Kawaguchi2008MNRAS,Kohri2008JCAP,Drees2011JCAP,Lin2013PLB,Linde2013PRD}. Other processes such as a sudden reduction in the pressure \cite{Khlopov1980PLB,Jedamzik1997PRD}, bubble collisions \cite{Crawford1982NATULONDON,Hawking1982PRD,Kodama1982PTP,La1989PLB,Moss1994PRD}, a curvaton \cite{Yokoyama1997AandA,Kawasaki2013PRD,Kohri2013PRD,Bugaev2013IJMPD} and collapse of cosmic string \cite{Hogan1984PLB} can lead to PBH production in the early Universe. For recent review, see \cite{Carr2010PRD,Carr2020ARNPS,Carr2020arXiv}.
 
In this paper, we study the effect of the scotogenic dark matter on the PBHs and vice versa. We show that if the PBHs evaporate in the radiation dominant era, the upper limit of the initial mass of the PBHs $M_{\rm in}$ should be constrained as $10^4 \lesssim M_{\rm in}/M_{\rm Pl} \lesssim  10^{10}$ for $\mathcal{O}$(1) TeV scotogenic dark matter ($\mathcal{O}$(1) TeV is most appropriate energy scale in the scotogenic model). On the other hand, if the PBHs evaporate in the PBH dominated era, a quite heavy scotogenic dark matter ($m_{\rm DM} \gtrsim 10^9$ GeV) for $M_{\rm in}/M_{\rm Pl} \sim 10^{13}$ may be allowed.

This paper is organized as follows. In Sec. \ref{sec:PBH} and Sec. \ref{sec:scotogenic}, we present reviews of the primordial black holes and the scotogenic model. Sec. \ref{sec:PBH} has been written mainly with references  
\cite{Fujita2014PRD,Morrison2019JCAP,Masina2020EPJP,Gondolo2020PRD,Bernal2020arXiv1}. In Sec. \ref{sec:PBH} and  Sec. \ref{sec:scotogenic}, we also show the notation and some basic assumptions in this paper. In Sec. \ref{sec:PBHsAndScotogenic}, we show the effect of the scotogenic dark matter produced by freeze-out mechanism on PBHs and vice versa. Section \ref{sec:summary} is devoted to a summary.

In this paper, we use the natural unit ($c=\hbar=k_{\rm B}=1$).

\section{Primordial black holes\label{sec:PBH}}

\subsection{Early Universe}
We show some fundamental knowledge of the early Universe. We note that our discussion is based on the cosmological principle, the Universe is homogeneous and isotropic, and we neglect the curvature and cosmological constant in the early Universe.

Evolution of the early Universe is described by the Friedmann equation 
\begin{eqnarray}
\left(\frac{\dot{a}(t)}{a(t)} \right)^2 = H(t)^2 = \frac{8\pi G}{3} \rho(t) = \frac{8\pi}{3M_{\rm Pl}^2}\rho(t),
\label{Eq:Friedmann}
\end{eqnarray}
where $a(t)$ is the scale factor, $H(t)$ is the Hubble parameter, $G \simeq 6.674 \times 10^{-11}$ m$^3$ kg$^{-1}$ s$^{-2}$ is the Newton gravitational constant, $\rho(t)$ is the energy density of the Universe and  $M_{\rm Pl}=\sqrt{1/G} \simeq 1.221 \times 10^{19}$ GeV is the Planck mass.

In the first stage of the early Universe, the main ingredients are the relativistic particles (these particles are called radiation). The energy density of radiation is given by
\begin{eqnarray}
\rho_{\rm rad}(T) = \frac{\pi^2}{30} g_*(T)T^4,
\label{Eq:rho_rad}
\end{eqnarray}
where
\begin{eqnarray}
g_*(T) = \sum_{i={\rm bosons}}g_i \left(\frac{T_i}{T}\right)^4+ \frac{7}{8} \sum_{i={\rm fermions}}g_i \left(\frac{T_i}{T}\right)^4,
\end{eqnarray}
denotes the relativistic effective degrees of freedom for the radiation energy density. In the radiation dominated era, we have $\rho=\rho_{\rm rad} \propto a^{-4}$, $a(t) \propto t^{1/2}$ and 
\begin{eqnarray}
H(t)=\frac{1}{2t}.
\end{eqnarray}
Eq. (\ref{Eq:Friedmann}) and Eq. (\ref{Eq:rho_rad}) yield 
\begin{eqnarray}
H(T) = \sqrt{\frac{4\pi^3 g_*(T)}{45}}\frac{T^2}{M_{\rm Pl}},
\end{eqnarray}
and we obtain the relation between temperature and time as
\begin{eqnarray}
T(t) = \left(\frac{45}{16\pi^3 g_*(T)} \right)^{1/4} \left( \frac{M_{\rm Pl}}{t} \right)^{1/2}.
\end{eqnarray}

The entropy density is given by
\begin{eqnarray}
s(T) = \frac{2\pi^2}{45} g_{*s}(T)T^3,
\label{Eq:s}
\end{eqnarray}
where
\begin{eqnarray}
g_{*s}(T) = \sum_{i={\rm bosons}}g_i \left(\frac{T_i}{T}\right)^3+ \frac{7}{8} \sum_{i={\rm fermions}}g_i \left(\frac{T_i}{T}\right)^3,
\end{eqnarray}
denotes the  relativistic effective degrees of freedom for the entropy density.

Since the energy density of the non-relativistic particles evolves as  $\rho_{\rm mat} \propto a^{-3}$, $a(t) \propto t^{2/3}$, after the radiation dominated era, non-relativistic particles (these particle are called matter) eventually dominate the energy density over radiation (recall that $\rho_{\rm rad} \propto a^{-4}$). 
In this matter dominated era, we have
\begin{eqnarray}
H(t)=\frac{2}{3t}.
\label{Eq:H=2/(3t)}
\end{eqnarray}
%

\subsection{PBH formation and evaporation}
\subsubsection{Formation}
We assume that PBHs are produced in the early Universe by large density perturbations generated from an inflation \cite{Garcia-Bellido1996PRD,Kawasaki1998PRD,Yokoyama1998PRD,Kawasaki2006PRD,Kawaguchi2008MNRAS,Kohri2008JCAP,Drees2011JCAP,Lin2013PLB,Linde2013PRD} and a PBH's mass is proportional to a horizon mass and that PBHs have the same masses at their formation time. Moreover, we assume that the PBHs form during the radiation dominated era, with a monochromatic mass function.

The initial mass of a PBH formed in a radiation dominated era is evaluated as 
\begin{eqnarray}
M_{\rm in} = M_{\rm BH}(T_{\rm in}) = \frac{4\pi}{3}\gamma \rho_{\rm rad} (T_{\rm in}) H(T_{\rm in})^{-3},
\label{Eq:PBHformationMin}
\end{eqnarray}
where $T_{\rm in}$ denotes the temperature of the Universe at PBH formation time, $\gamma$ is a numerical factor that depends on the details of the gravitational collapse. According to Carr's formula, $\gamma \sim 0.2$ \cite{Carr1975APJ}. The temperature of the Universe at PBH formation time is obtained as 
\begin{eqnarray}
T_{\rm in} = \frac{\sqrt{3}5^{1/4}}{2\pi^{3/4}}\frac{\gamma^{1/2}}{g_*(T_{\rm in})^{1/4}}\left( \frac{M_{\rm Pl}^3}{M_{\rm in}} \right)^{1/2},
\label{Eq:Tin}
\end{eqnarray}
by Eqs. (\ref{Eq:Friedmann}), (\ref{Eq:rho_rad}) and (\ref{Eq:PBHformationMin}).

We introduce the dimensionless parameter
\begin{eqnarray}
\beta = \frac{\rho_{\rm BH}(T_{\rm in})}{\rho_{\rm rad}(T_{\rm in})} = \frac{M_{\rm in} n_{\rm BH}(T_{\rm in})}{\rho_{\rm rad}(T_{\rm in})},
\label{Eq:beta}
\end{eqnarray}
where  $\rho_{\rm BH}$ is the density of the PBHs, to represent the initial energy density of PBHs at the time of its formation. 

\subsubsection{Evaporation}
A black hole loses its mass by producing particles with masses below the Hawking temperature (horizon temperature of the black hole)
\begin{eqnarray}
T_{\rm BH} = \frac{M_{\rm Pl}^2}{8\pi M_{\rm BH}},
\end{eqnarray}
via Hawking radiation \cite{Hawking1975CMP}. Ignoring gray body factors, the Hawking radiation can be described as black body radiation. The energy spectrum of the Hawking radiation is similar to the Planck distribution
\begin{eqnarray}
\frac{d^2 u_i (E,t)}{dt dE} = \frac{g_i}{8\pi^2} \frac{E^3}{e^{E/T_{\rm BH}}\pm 1},
\label{Eq:d2u_dtdE}
\end{eqnarray}
where $u_i$ is the total radiated energy per unit area, $g_i$ is the degrees of  freedom of the $i$th species being radiated, $E$ is the total energy of the emitted particles, $+$ for fermion emission and $-$ for boson emission. We note that the Hawking radiation Eq. (\ref{Eq:d2u_dtdE}) differs by a factor of 27/4 with that including the gray body factor in the high energy geometrical optics limit \cite{Carr2010PRD,Lennon2018JCAP,Baldes2020JCAP,Auffinger2020arXiv}.

The time evolution of the black hole mass due to Hawking radiation is given by
\begin{eqnarray}
\frac{dM_{\rm BH}(t)}{dt} = -4\pi r_S^2 \sum_i \int_0^\infty \frac{d^2 u_i (E,t)}{dt dE} dE  
 = -\frac{g_*(T_{\rm BH})}{30720\pi}\frac{M_{\rm Pl}^4}{M_{\rm BH}(t)^2},
\end{eqnarray}
where $r_s = 2M_{\rm BH}/M_{\rm Pl}^2$ is the Schwarzschild radius of the black hole, $g_*(T_{\rm BH})$ is the total number of relativistic degrees of freedom emitted by the black hole. If $g_*(T_{\rm BH})$ is constant, the time evolution of the mass of a black hole with initial mass $M_{\rm in}$ formed at $t_{\rm in}$ is obtained as
\begin{eqnarray}
M_{\rm BH}(t) = M_{\rm in} \left( 1 - \frac{t-t_{\rm in}}{\tau} \right)^{1/3},
\end{eqnarray}
where
\begin{eqnarray}
\tau = \frac{10240\pi}{g_* (T_{\rm BH})}\frac{M_{\rm in}^3}{M_{\rm Pl}^4},
\label{Eq:tau}
\end{eqnarray}
is the lifetime of the black hole. Since
\begin{eqnarray}
\frac{t_{\rm in}}{\tau} = \frac{g_*(T_{\rm BH})}{10240 \pi \gamma} \left( \frac{M_{\rm Pl}}{M_{\rm in}} \right)^2 \ll 1,
\end{eqnarray}
for appropriate initial mass of PBHs, $10^4 \le M_{\rm in}/M_{\rm Pl} \le 10^{13}$ (we see later), the time of evaporation of PBH $t_{\rm evap} $ is almost same as the lifetime of the PBH, $t_{\rm evap} = t_{\rm in} + \tau \sim \tau$. Thus, the temperature of the Universe right after PBH evaporation is
\begin{eqnarray}
T_{\rm evap} &=& T(t_{\rm in} + \tau) \sim T(\tau) 
= \frac{\sqrt{3} g_* (T_{\rm BH})^{1/4}}{64 \sqrt{2}5^{1/4}\pi^{5/4}} \left( \frac{M_{\rm Pl}^5}{M_{\rm in}^3}\right)^{1/2} \nonumber \\
&=& 1.20 \times 10^{17} \left(\frac{g_* (T_{\rm BH})}{106.75}\right)^{1/4}  \left( \frac{M_{\rm Pl}}{M_{\rm in}}\right)^{3/2}.
\label{Eq:Tevap}
\end{eqnarray}

The rate of $i$th species emission is expressed as 
\begin{eqnarray}
\frac{d^2 N_i(E,T)}{dtdE} = \frac{4\pi r_S^2}{E}\frac{d^2u_i(E,t)}{dt dE}.
\end{eqnarray}
The total number of the $i$th species with mass $m_i$ emitted from a single black hole is
\begin{eqnarray}
N_i = \int_{t(m_i)}^{t_{\rm evap} \sim \tau} dt \int_0^\infty \frac{d^2 N_i(E,T)}{dtdE} dE,
\label{Eq:intN_i}
\end{eqnarray}
with
\begin{eqnarray}
t(m_i) = 
\begin{cases}
t_{\rm in}  & {\rm for} \ T_{\rm BH}^{\rm in} > m_i \\
t(T_{\rm BH} = m_i) &  {\rm for} \ T_{\rm BH}^{\rm in} < m_i \
\end{cases},
\end{eqnarray}
where
\begin{eqnarray}
t(T_{\rm BH} = m_i) = t_{\rm in} + \tau\left[ 1 - \left( \frac{1}{8\pi m_i} \frac{M_{\rm Pl}^2 }{M_{\rm in}}\right)^3 \right],
\end{eqnarray}
denotes the time as which black hole start emitting $i$th species. By integrating Eq. (\ref{Eq:intN_i}), we obtain
\begin{eqnarray}
N_i &=& \frac{120 \zeta(3)}{\pi^3} \frac{g_i C_i}{g_*(T_{\rm BH})} \left(\frac{M_{\rm in}}{M_{\rm Pl}} \right)^2 \  {\rm for} \ T_{\rm BH}^{\rm in} > m_i, \nonumber \\
N_i &=&\frac{15 \zeta(3)}{8\pi^5} \frac{g_i C_i}{g_*(T_{\rm BH})} \left(\frac{M_{\rm Pl}}{m_i} \right)^2 \  {\rm for} \ T_{\rm BH}^{\rm in} < m_i, 
\label{Eq:Ni}
\end{eqnarray}
where
\begin{eqnarray}
C_i = 
\begin{cases}
1  & {\rm for} \ {\rm bosons} \\
3/4 &  {\rm for} \ {\rm fermions}\
\end{cases},
\end{eqnarray}
and
\begin{eqnarray}
T_{\rm BH}^{\rm in} = \frac{M_{\rm Pl}^2}{8\pi M_{\rm in}} = 4.86 \times 10^{17} \ {\rm GeV} \times \frac{M_{\rm Pl}}{M_{\rm in}}.
\label{Eq:PBH_Tin_MinPMpl}
\end{eqnarray}

The mean energy of the emitted particle is \cite{Bernal2020arXiv1,Fujita2014PRD,Morrison2019JCAP,Baumann2007arXiv} 
\begin{eqnarray}
\braket{E_i} &=& \frac{1}{N_i}\int_{t(m_i)}^{t_{\rm evap}}dt\int_0^\infty dE 3T_{\rm BH} \frac{d^2 N_i}{dt dE}
= 6 \times
\begin{cases}
T_{\rm BH}^{\rm in} &  \ {\rm for} \ T_{\rm BH}^{\rm in} > m_i \\
m_{\rm DM} & \ {\rm for} \ T_{\rm BH}^{\rm in} < m_i 
\end{cases},
\label{Eq:<E>}
\end{eqnarray}
where $3T_{\rm BH}$ is the average energy of particles radiated by a PBH with temperature $T_{\rm BH}$.

\subsection{Dark matter production by PBH}
In this paper, we use the following numerical values for the critical density
\begin{eqnarray}
\rho_{\rm c} &=& \frac{3H_0^2}{8\pi G} = 1.0537 \times 10^{-5} h^2 \ {\rm GeV} \ {\rm cm}^{-3} 
= 8.0964 \times 10^{-47} h^2 \ {\rm GeV}^4,
\label{Eq:rho_c}
\end{eqnarray}
and for the entropy density today
\begin{eqnarray}
s(T_0) = 2891.2 \left( \frac{T_0}{2.7255} \right)^3 {\rm cm}^{-3},
\label{Eq:s_0}
\end{eqnarray}
where $H_0= 100 h  \ {\rm km \ s^{-1} Mpc^{-1}}$ is the Hubble parameter today, $h$ is the dimension less Hubble parameter today, and $T_0$ is the temperature of the radiation today.

\subsubsection{PBH evaporation in radiation dominated era}
If all PBHs have evaporated in radiation dominated era, the total number density of dark matter particles emitted from all PBHs in the early Universe is obtained as $n_{\rm DM} = N_{\rm DM} n_{\rm BH}(T_{\rm in})$ where $n_{\rm BH}(T_{\rm in})$ denotes the initial number density of the PBHs. In terms of the comoving density $Y_{\rm DM}= n_{\rm DM}/s$, the total number density of dark matter produced by Hawking evaporation of PBHs in a radiation dominated era can be estimated by
\begin{eqnarray}
Y_{\rm DM}(T_0) = \frac{n_{\rm DM}(T_0)}{s(T_0)} 
 = \frac{n_{\rm DM}(T_{\rm evap})}{s(T_{\rm evap})}  
 = \frac{N_{\rm DM}n_{\rm BH}(T_{\rm in})}{s(T_{\rm in})},
 \label{Eq:YDM}
\end{eqnarray}
where the conservation of the entropy of the standard model particles is used. By combining Eqs. (\ref{Eq:rho_rad}), (\ref{Eq:s}), (\ref{Eq:Tin}), (\ref{Eq:beta}) and (\ref{Eq:YDM}), we obtain 
\begin{eqnarray}
Y_{\rm DM} (T_0)&=&  \frac{3}{4} N_{\rm DM} \beta \frac{g_*(T_{\rm BH}) }{g_{*s}(T_{\rm BH}) }\frac{T_{\rm in}}{M_{\rm in}} \\
 &\simeq& \frac{3\sqrt{3} 5^{1/4} g_* (T_{\rm in})^{-1/4}\beta \gamma^{1/2}}{8 \pi^{3/4}}  N_{\rm DM} \left( \frac{M_{\rm Pl}}{M_{\rm in}} \right)^{3/2}, \nonumber 
\end{eqnarray}
where we use $g_*(T_{\rm BH})  \simeq g_{*s}(T_{\rm BH})$ to obtain the second expression. 

The relic abundance of dark matter which is generated by PBH evaporation can be expressed in term of the density parameter as 
\begin{eqnarray}
\Omega_{\rm PBH} =  \frac{\rho_{\rm DM}(T_0)}{\rho_{\rm c}} = \frac{m_{\rm DM} Y_{\rm DM}(T_0)s(T_0)}{\rho_{\rm c}}.
\label{Eq:OmegaDM}
\end{eqnarray}
By inserting numerical values of $\rho_{\rm c}$ and $s(T_0)$ into Eq. (\ref{Eq:OmegaDM}), we obtain
\begin{eqnarray}
\Omega_{\rm PBH}  h^2 &\simeq&  7.31 \times 10^7  \left( \frac{\gamma}{0.2} \right)^{1/2} \left(\frac{g_*(T_{\rm in})}{106.75} \right)^{-1/4}   \label{Eq:OmegaPBHh2_RD_T>m} \\
 &&  \times  \beta C_{\rm DM} \frac{g_{\rm DM}  }{g_*(T_{\rm BH})}   \left( \frac{m_{\rm DM}}{{\rm GeV}} \right) \left( \frac{M_{\rm in}}{M_{\rm Pl}} \right)^{1/2}, \nonumber
\end{eqnarray}
for $T_{\rm BH}^{\rm in} > m_{\rm DM}$ and
\begin{eqnarray}
\Omega_{\rm PBH}  h^2 &\simeq&  1.16 \times 10^5  \left( \frac{\gamma}{0.2} \right)^{1/2} \left(\frac{g_*(T_{\rm in})}{106.75} \right)^{-1/4}  \label{Eq:OmegaPBHh2_RD_T<m} \\
 &&  \times  \beta C_{\rm DM} \frac{g_{\rm DM}  }{g_*(T_{\rm BH})}   \left( \frac{m_{\rm DM}}{{\rm GeV}} \right) \left( \frac{M_{\rm Pl}^7}{M_{\rm in}^3 m_{\rm DM}^4} \right)^{1/2},
\nonumber 
\end{eqnarray}
for $T_{\rm BH}^{\rm in} < m_{\rm DM}$.

Since $\rho_{\rm PBH} \propto a^{-3}$ and $\rho_{\rm rad}\propto a^{-4}$, $\rho_{\rm PBH}(t_{\rm early-eq}) \simeq \rho_{\rm rad}(t_{\rm early-eq})$ may be happen at the early equality time $t_{\rm early-eq}$ defined by 
\begin{eqnarray}
\frac{\rho_{\rm PBH}(t_{\rm early-eq})}{\rho_{\rm rad}(t_{\rm early-eq})} = \frac{\rho_{\rm PBH}(T_{\rm early-eq})}{\rho_{\rm rad}(T_{\rm early-eq})} 
= \frac{\rho_{\rm PBH}(T_{\rm in})}{\rho_{\rm rad}(T_{\rm in})}\frac{T_{\rm in}}{T_{\rm early-eq}} 
= \beta_{\rm c}\frac{T_{\rm in}}{T_{\rm early-eq}} \simeq 1. \nonumber \\
\end{eqnarray}
In order for PBH evaporation to occur before the early equality time, $t_{\rm evap} \lesssim t_{\rm early-eq}$ or equivalently to happen in the radiation dominated (RD) era,  the condition  
\begin{eqnarray}
\beta  < \beta_{\rm c} &=& \frac{T_{\rm eva}}{T_{\rm in}} = \sqrt{\frac{g_*(T_{\rm BH})}{10240\pi\gamma}}\left(\frac{M_{\rm in}} {M_{\rm Pl}}\right)^{-1} \label{Eq:RDconstraint} \\
&=& 0.129 \left(\frac{g_*(T_{\rm BH})}{106.75}\right)^{1/2} \left(\frac{0.2}{\gamma}\right)^{1/2}\left(\frac{M_{\rm in}} {M_{\rm Pl}}\right)^{-1},\nonumber
\end{eqnarray}
should be satisfied (RD constraint) \cite{Fujita2014PRD,Morrison2019JCAP,Masina2020EPJP,Gondolo2020PRD,Bernal2020arXiv1}.

\subsubsection{PBH evaporation in matter dominated era}
If PBH evaporation happens after the early equality time, $t_{\rm early-eq} \lesssim t_{\rm evap}$, or equivalently, the condition  
\begin{eqnarray}
\beta  > \beta_{\rm c},
\end{eqnarray}
is satisfied, the PBH evaporation happens in the matter dominated era (PBH dominated era). 

The total number density of dark matter produced by Hawking evaporation of PBHs is evaluated as \cite{Bernal2020arXiv1}
\begin{eqnarray}
Y_{\rm DM}(T_0) = \frac{n_{\rm DM}(T_0)}{s(T_0)} 
= \frac{n_{\rm DM}(\tilde{T}_{\rm evap})}{s(\tilde{T}_{\rm evap})} 
= \frac{N_{\rm DM}n_{\rm BH}(t_{\rm evap})}{s(\tilde{T}_{\rm evap})},
\end{eqnarray}
where $\tilde{T}_{\rm evap}$ is the temperature that the radiated particles from PBHs equilibrate to the radiation of the standard model particles just after the complete black hole evaporation, and the conservation of the entropy of the standard model particle after the PBHs have completely evaporated is used. Assuming an instantaneous evaporation of the PBHs at $t = t_{\rm evap} \simeq \tau$, we have 
\begin{eqnarray}
n_{\rm BH}(t_{\rm evap}) = \frac{\rho_{\rm BH}(t_{\rm evap})}{M_{\rm in}} \simeq \frac{\rho_{\rm BH}(\tau)}{M_{\rm in}}.
\end{eqnarray}
By combining Eqs. (\ref{Eq:Friedmann}), (\ref{Eq:H=2/(3t)}) and (\ref{Eq:tau}), we obtain the energy density of the black hole in the PBH dominated era as
\begin{eqnarray}
\rho_{\rm BH}(\tau) = \frac{M_{\rm Pl}^2}{6\pi\tau^2} = \frac{M_{\rm Pl}^2}{6\pi} \left(\frac{g_* (T_{\rm BH})}{10240\pi} \frac{M_{\rm Pl}^4}{M_{\rm in}^3} \right)^2.
\end{eqnarray}
Since the density of the PBH just after the complete black hole evaporation is approximately 
\begin{eqnarray}
\rho_{\rm BH}(\tau)  \simeq \frac{\pi^2}{30}g_* (T_{\rm BH}) \tilde{T}_{\rm evap}^4,
\end{eqnarray}
we obtain
\begin{eqnarray}
\tilde{T}_{\rm evap} \simeq  \frac{g_* (T_{\rm BH})^{1/4}}{32 \sqrt{2} 5^{1/4}\pi^{5/4}}   \left( \frac{M_{\rm Pl}^{5}}{M_{\rm in}^3} \right)^{1/2},
\end{eqnarray}
and 
\begin{eqnarray}
Y_{\rm DM} (T_0) \simeq \frac{3}{4} N_{\rm DM} \frac{g_*(T_{\rm BH}) }{g_{*s}(T_{\rm BH}) }\frac{\tilde{T}_{\rm evap}}{M_{\rm in}} 
 \simeq \frac{3g_* (T_{\rm BH})^{1/4}}{128 \sqrt{2} 5^{1/4}\pi^{5/4}}  N_{\rm DM} \left( \frac{M_{\rm Pl}}{M_{\rm in}} \right)^{5/2},
\end{eqnarray}
where we use $g_*(T_{\rm BH})  \simeq g_{*s}(T_{\rm BH})$ to obtain the second expression. 

In term of the density parameter, the relic abundance of dark matter which is generated by PBH evaporation is  
\begin{eqnarray}
\Omega_{\rm PBH}  h^2 &\simeq&  1.09 \times 10^7  \left(\frac{g_*(T_{\rm BH})}{106.75} \right)^{1/4} \label{Eq:OmegaPBHh2_MD_T>m} \\
 &&  \times  C_{\rm DM} \frac{g_{\rm DM}  }{g_*(T_{\rm BH})}   \left( \frac{m_{\rm DM}}{{\rm GeV}} \right) \left( \frac{M_{\rm Pl}}{M_{\rm in}} \right)^{1/2},
\nonumber 
\end{eqnarray}
for $T_{\rm BH}^{\rm in} > m_{\rm DM}$ or
\begin{eqnarray}
\Omega_{\rm PBH}  h^2 &\simeq&  1.72 \times 10^4  \left(\frac{g_*(T_{\rm BH})}{106.75} \right)^{1/4} \label{Eq:OmegaPBHh2_MD_T<m} \\
 &&  \times  C_{\rm DM} \frac{g_{\rm DM}  }{g_*(T_{\rm BH})}   \left( \frac{m_{\rm DM}}{{\rm GeV}} \right) \left( \frac{M_{\rm Pl}^9}{M_{\rm in}^5 m_{\rm DM}^4} \right)^{1/2},
\nonumber 
\end{eqnarray}
for $T_{\rm BH}^{\rm in} < m_{\rm DM}$

\subsection{Constraints from CMB, BBN and WDM}
\subsubsection{CMB and BBN}
Since the Hubble parameter $H(t)$ at time $t$ is less than or equal to the Hubble parameter during inflation, we obtain the lower limit of the initial mass of PBH as $M_{\rm in} \gtrsim 0.1$ g  (cosmic microwave background radiation (CMB) constraint) \cite{Fujita2014PRD}. Moreover, PBHs should be evaporated before big bang nucleosynthesis, the upper limit of the initial mass $M_{\rm in} < 10^9$ g is obtained (BBN constraint). Thus, we obtain $0.1 {\rm g} \le M_{\rm in} \le 10^9 {\rm g}$ as well as $4.6 \times 10^3 \lesssim M_{\rm in}/M_{\rm Pl} \lesssim 4.6 \times 10^{13}$ from CMB and BBN constraints. In this paper we require conservatively 
\begin{eqnarray}
10^4 \le \frac{M_{\rm in}}{M_{\rm Pl}} \le  10^{13},
\label{Eq:CMB_BBN_constraints}
\end{eqnarray}
as the CMB and BBN constraints.

\subsubsection{WDM}
The dark matter should be cold enough to avoid erasing small-scale structures via free streaming. Thus the dark matter should be warm dark matter (WDM) or should be more slow dark matter (cold dark matter). If the dark matter has no interaction with any particles in the thermal bath, the only way that dark matter particles can lose energy (or momentum) is by redshifting. 

The redshifted present momentum of a dark matter particle $p(t_0)$ is  
\begin{eqnarray}
p(t_0) = \frac{a(t_{\rm evap})}{a(t_0)} p(t_{\rm evap}) 
= \frac{a(t_{\rm evap})}{a(t_{\rm eq})} \frac{a(t_{\rm eq})}{a(t_0)}   p(t_{\rm evap}) 
= \frac{a(t_{\rm evap})}{a(t_{\rm eq})} \frac{\Omega_{\rm rad}}{\Omega_{\rm mat}}   p(t_{\rm evap}).
\end{eqnarray}
By using the relation $\rho_{\rm rad} \propto a^{-4}$, we have
\begin{eqnarray}
 \frac{a(t_{\rm evap})}{a(t_{\rm eq})} 
= \left(\frac{\rho_{\rm rad}(T_{\rm eq})}{\rho_{\rm rad}(T_{\rm evap})}\right)^{1/4}.
\end{eqnarray}
According to the following relations
\begin{eqnarray}
\rho_{\rm rad}(T_{\rm eq}) &=& \rho_{\rm c} \left(\frac{a(T_0)}{a(T_{\rm eq})}\right)^3= \rho_{\rm c} a(T_{\rm eq})^{-3}, \quad
\rho_{\rm rad}(T_{\rm evap}) \simeq \frac{3M_{\rm Pl}^2}{8\pi}\frac{1}{4\tau^2}, \nonumber \\ 
a(T_{\rm eq}) &=& a(t_{\rm eq}) =  \frac{\Omega_{\rm rad}}{\Omega_{\rm mat}}, 
\end{eqnarray}
and $p(t_{\rm evap}) \simeq \braket{E} \simeq 6 T_{\rm BH}^{\rm in}$ for light dark matter ($m_{\rm DM} \ll T_{\rm BH}^{\rm in}$), momentum of a dark matter particle today is obtained as
\begin{eqnarray}
p(t_0)= \frac{16\sqrt{5}6^{3/4}\rho_c^{1/4}}{\pi^{1/4}\sqrt{ g_*(T_{\rm BH})}} \left(\frac{\Omega_{\rm rad}}{\Omega_{\rm mat}} \right)^{1/4} \left( \frac{M_{\rm in}}{M_{\rm Pl}}\right)^{1/2}.
\end{eqnarray}
For $g_*(T_{\rm BH}) = 106.75$, $\Omega_{\rm rad} = 5.38\times 10^{-5}$, $\Omega_{\rm mat} = 0.315$ \cite{PDG}, we have
\begin{eqnarray}
\left(\frac{p(t_0)}{{\rm GeV}}\right) = 3.42 \times 10^{-12} \left( \frac{M_{\rm in}}{M_{\rm Pl}}\right)^{1/2}. 
\end{eqnarray}
Taking the warm dark matter to be heavier than 3.5 keV \cite{Irsic2017PRD}, the upper limit of the dark matter velocity is estimated as $v_{\rm DM} \lesssim 1.8 \times 10^{-8}$ \cite{Masina2020EPJP}. From $p(t_0) = m_{\rm DM} v_{\rm DM}$, we obtain the following constraint on mass of dark matter and initial mass of PBH
\begin{eqnarray}
\left( \frac{m_{\rm DM}}{{\rm GeV}} \right) \gtrsim 1.90 \times 10^{-4} \left( \frac{M_{\rm in}}{M_{\rm Pl}}\right)^{1/2},
\label{Eq:mDMlowerFromWDM}
\end{eqnarray}
as the WDM constraint \cite{Fujita2014PRD}.

We note that the WDM constraint only applies if all of dark matter has the large velocity dispersion. In the warm-plus-cold dark case, there is a slow dark matter component and the WDM constraint will be weaker \cite{Baldes2020JCAP,Boyarsky2009JCAP,Baur2017JCAP}.   

\section{Scotogenic model\label{sec:scotogenic}}
\subsection{Model}
The scotogenic model \cite{Ma2006PRD} is an extension of the standard model in the particle physics. In this model,  three new Majorana $SU(2)_L$ singlets $N_k$ $(k=1,2,3)$ and one new scalar $SU(2)_L$ doublet $(\eta^+,\eta^0)$ are introduced. These new particles are odd under exact $Z_2$ symmetry. Under $SU(2)_L \times U(1)_Y \times Z_2$, the main particle contents in the scotogenic model is given by  $(\alpha=e,\mu,\tau)$ :
\begin{eqnarray}
&& L_\alpha=(\nu_\alpha, \ell_\alpha)_L \ : \ (2,-1/2,+), \quad \ell_\alpha^C \ : \ (1,1,+),\nonumber \\
&& \Phi=(\phi^+, \phi^0) \ : \ (2,1/2,+), \nonumber \\
&& N_k \ : \  (1,0,-), \quad \eta=(\eta^+,\eta^0) \ : \ (2,1/2,-),  
\end{eqnarray}
where $(\nu_\alpha, \ell_\alpha)$ is the left-handed lepton doublet and $(\phi^+, \phi^0)$ is the standard Higgs doublet. 

The relevant Lagrangian and scalar potential for this paper are given by
\begin{eqnarray}
\mathcal{L} &=& Y_{\alpha k} (\bar{\nu}_{\alpha L} \eta^0 - \bar{\ell}_{\alpha L} \eta^+) N_k + \frac{1}{2}M_k \bar{N}_k N^C_k + h.c.,
\nonumber \\
V &=& \frac{1}{2}\lambda (\Phi^\dagger \eta)^2 + h.c.,
\label{Eq:L_V}
\end{eqnarray}
where $Y$ is the Yukawa matrix, $M_k$ is the mass of $N_k$ and $\lambda$ is the coupling constant for the $\Phi\Phi\eta\eta$ interaction.

Owing to the $Z_2$ symmetry, the tree level neutrino mass should vanish but they acquire masses via one-loop interactions,
\begin{eqnarray}
M_{\alpha\beta} = \sum_{k=1}^3 Y_{\alpha k}Y_{\beta k} \Lambda_k,
\label{Eq:M_alpha_beta}
\end{eqnarray}
where
\begin{eqnarray}
\Lambda_k =  \frac{\lambda v^2}{16\pi^2}\frac{M_k}{m^2_0-M^2_k}\left(1-\frac{M^2_k}{m^2_0-M^2_k}\ln\frac{m_0^2}{M^2_k} \right),
\quad
m_0^2 = \frac{1}{2}(m_R^2+m_I^2),
v\label{Eq:Lambda_k} 
\end{eqnarray}
and $v$, $m_R$, $m_I$ denote vacuum expectation value of the Higgs field, the masses of $\sqrt{2} {\rm Re}[\eta^0]$ and $\sqrt{2} {\rm Im}[\eta^0]$, respectively. 

In this model, flavor violating processes such as $\mu \rightarrow e \gamma$ are induced at the one-loop level. The branching ratio of $\mu \rightarrow e \gamma$  is given by \cite{Kubo2006PLB}
\begin{eqnarray}
{\rm BR}(\mu \rightarrow e \gamma)=\frac{3\alpha_{\rm em}}{64\pi(G_{\rm F} m_0^2)^2}\left| \sum_{k=1}^3 Y_{\mu k}Y_{e k}^* F \left( \frac{M_k}{m_0}\right) \right|^2,
\nonumber \\
\end{eqnarray}
where $\alpha_{\rm em}$ denotes the fine-structure constant, $G_{\rm F}$ denotes the Fermi coupling constant and  $F(x)$ is defined by
\begin{eqnarray}
F(x)=\frac{1-6x^2+3x^4+2x^6-6x^4 \ln x^2}{6(1-x^2)^4}.
\end{eqnarray}
%

\subsection{Yukawa matrix and neutrino parameters}
In order to  obtain any phenomenological prediction in the scotogenic model, the elements of the Yukawa matrix $Y$ should be determined. The Yukawa matrix  can be written using by so-called Casas-Ibarra parametrization \cite{Casas2001NPB,Toma2014JHEP}
\begin{eqnarray}
Y = D_{\sqrt{\Lambda^{-1}}} R D_{\sqrt{M_\nu^{\rm diag}}}U^\dag,
\end{eqnarray}
where
\begin{eqnarray}
D_{\sqrt{\Lambda^{-1}}} &=& {\rm diag.}(\sqrt{\Lambda_1},\sqrt{\Lambda_2},\sqrt{\Lambda_3}), \nonumber \\
D_{\sqrt{M_\nu^{\rm diag}}} &=& {\rm diag.}(\sqrt{m_1},\sqrt{m_2},\sqrt{m_3}),
\end{eqnarray}
and $R$ is an arbitrary $3 \times 3$ complex orthogonal matrix which satisfies $R^TR=I$ where $I$ is the unit matrix. The neutrino mass eigenvalue $m_i$, $(i=1,2,3)$ is related to the flavor neutrino mass matrix $M_\nu$:
\begin{eqnarray}
 U^T M_\nu U = M_\nu^{\rm diag} = {\rm diag.}(m_1,m_2,m_3), 
\end{eqnarray}
where 
\begin{eqnarray}
M_\nu =  \left( 
\begin{array}{ccc}
M_{ee} & M_{e\mu}  &  M_{e\tau}\\
 -  & M_{\mu\mu}  &  M_{\mu\tau}\\
-  & -  &  M_{\tau\tau} \\
\end{array}
\right),
\end{eqnarray}
with the symbol ``-" denotes a symmetric partner, and
\begin{eqnarray}
U=
 \left(
\begin{matrix}
c_{12}c_{13} & s_{12}c_{13} & \tilde{s}_{13}\\
- s_{12}c_{23} - c_{12}\tilde{s}_{13}^*s_{23} & c_{12}c_{23} - s_{12}\tilde{s}_{13}^*s_{23} & c_{13}s_{23}\\
s_{12}s_{23} - c_{12}\tilde{s}_{13}^*c_{23} & - c_{12}s_{23} - s_{12}\tilde{s}_{13}^*c_{23} & c_{13}c_{23}
\end{matrix}
\right) 
 \left(
\begin{matrix}
e^{i\alpha_1} & 0 & 0\\
0 & e^{i\alpha_2} & 0\\
0 & 0 & 1
\end{matrix}
\right),
\label{Eq:UPMNS}
\end{eqnarray}
is the neutrino mixing matrix (Pontecorvo-Maki-Nakagawa-Sakata (PMNS) matrix) \cite{Pontecorvo1957,Pontecorvo1958,Maki1962PTP}. We employ the PDG parametrization of the PMNS martix \cite{PDG}. We use the abbreviations $c_{ij}=\cos\theta_{ij}$, $s_{ij}=\sin\theta_{ij}$  ($i,j$=1,2,3) and $\tilde{s}_{13}=s_{13}e^{-i\delta}$ where $\theta_{ij}$ is a neutrino mixing angle. The Dirac CP phase is denoted by $\delta$ and the Majorana CP phases are denoted by $\alpha_1$ and $\alpha_2$. In this paper, we assume that the mass matrix of the charged leptons is diagonal and real.

\subsection{Dark matter production by freeze-out}
Since the lightest $Z_2$ odd particle is stable, the lightest $Z_2$ odd particle becomes a dark matter candidate. In this paper, we assume that the lightest Majorana singlet fermion, $N_1$, is the dark matter particle. Moreover, we assume that  $N_1$ is almost degenerate with the next to lightest Majorana singlet fermion $N_2$ ($M_1 \lesssim M_2 < M_3$) to take account of coannihilation effects \cite{Griest1991PRD}. If we take account the coannihilation effect, the predicted cold dark matter abundance as well as the branching ratio of lepton flavor violating $\mu \rightarrow e \gamma$ process can be simultaneously consistent with observations within the scotogenic model \cite{Suematsu2009PRD,Suematsu2010PRD}.

The lightest $Z_2$ odd particles (dark matter particles) $N_1$ can reach thermal equilibrium with the thermal bath particles $f$ in the early Universe through annihilation and pair production processes ($N_1 N_1 \leftrightarrow \bar{f} f$). When the temperature of the bath particles drops below the mass of the dark matter particles, the equilibrium dark matter relic abundance is suppressed exponentially by Boltzmann factor until the $\Gamma \lesssim H$, where $\Gamma$ is the reaction rate of the annihilation $N_1 N_1 \rightarrow \bar{f} f$. Then, the comoving number density of dark matter particles remains constant. This is the so-called freeze-out scenario.  The freeze-out happens when \cite{KolbTurner1991}
\begin{eqnarray}
x_{\rm FO} = \frac{m_{\rm DM}}{T_{\rm FO}} \simeq 25,
\label{Eq:xf25}
\end{eqnarray}
where $m_{\rm DM}$ is the dark matter mass and $T_{\rm FO}$ is the freeze-out temperature.

In the freeze-out scenario, relic abundance of dark matter particles is given by 
\begin{eqnarray}
\Omega_{\rm FO} \propto \frac{1}{\langle \sigma_{\rm eff}|v_{\rm rel}| \rangle},
\end{eqnarray}
where $\langle \sigma_{\rm eff}|v_{\rm rel}| \rangle$ is the thermally averaged cross section for $N_1 N_1 \rightarrow \bar{f} f$ process. The (co)annihilation cross section times the relative velocity of annihilation particles $v_{\rm rel}$ is given by \cite{Suematsu2009PRD}
\begin{eqnarray}
\sigma_{ij} |v_{\rm rel}|= a_{ij} + b_{ij} v_{\rm rel}^2,
\end{eqnarray}
with
\begin{eqnarray}
a_{ij}&=& \frac{1}{8\pi}\frac{M_1^2}{(M_1^2+m_0^2)^2} \sum_{\alpha\beta}(Y_{\alpha i} Y_{\beta j} - Y_{\alpha j} Y_{\beta i})^2, \nonumber \\
b_{ij}&=&\frac{m_0^4-3m_0^2M_1^2-M_1^4}{3(M_1^2+m_0^2)^2}a_{ij} 
 +  \frac{1}{12\pi}\frac{M_1^2(M_1^4+m_0^4)}{(M_1^2+m_0^2)^4}  \sum_{\alpha\beta}Y_{\alpha i} Y_{\alpha j} Y_{\beta i} Y_{\beta j},
\label{Eq:a_b}
\end{eqnarray}
where $\sigma_{ij}$ $(i,j=1,2)$ is annihilation cross section for $N_i N_j \rightarrow \bar{f}f$. The effective cross section $\sigma_{\rm eff}$ is obtained as \cite{Griest1991PRD}
\begin{eqnarray}
\sigma_{\rm eff} &=& \frac{g_1^2}{g_{\rm eff}^2}\sigma_{11} + \frac{2g_1g_2}{g_{\rm eff}^2}\sigma_{12} (1+\Delta M)^{3/2}e^{-\Delta M \cdot x}+ \frac{g_2^2}{g_{\rm eff}^2}\sigma_{22} (1+\Delta M)^3 e^{-2\Delta M \cdot x},\nonumber \\
g_{\rm eff}&=&g_1+g_2 (1+\Delta M)^{3/2}e^{-\Delta M \cdot x}.
\end{eqnarray}
where $\Delta M = (M_2-M_1)/M_1$ depicts the mass splitting ratio of the degenerate singlet fermions, $x = M_1/T$ denotes the ratio of the mass of lightest singlet fermion to the temperature $T$ and $g_1$ and $g_2$ are the number of degrees of freedom of $N_1$ and $N_2$, respectively. Since $M_1 \simeq M_2$, we have $\Delta M \simeq 0$ and obtain
\begin{eqnarray}
\sigma_{\rm eff} |v_{\rm rel}|&=& \left(\frac{\sigma_{11}}{4} + \frac{\sigma_{12}}{2} + \frac{\sigma_{22}}{4}\right) |v_{\rm rel}| 
= a_{\rm eff} + b_{\rm eff} v_{\rm rel}^2,
\end{eqnarray}
where
\begin{eqnarray}
a_{\rm eff}= \frac{a_{11}}{4}+\frac{a_{12}}{2}+\frac{a_{22}}{4},  \quad
b_{\rm eff}= \frac{b_{11}}{4}+\frac{b_{12}}{2}+\frac{b_{22}}{4}.
\label{Eq:aeff_beff}
\end{eqnarray}

The thermally averaged cross section can be written as $\langle \sigma_{\rm eff}|v_{\rm rel}| \rangle = a_{\rm eff} + 6b_{\rm eff}/x$ and the relic abundance of cold dark matter which is produced by freeze-out mechanism is estimated to be:
\begin{eqnarray}
\Omega_{\rm FO} h^2 = \frac{1.07\times 10^9 x_{\rm FO}}{g_\ast^{1/2} M_{\rm Pl} (a_{\rm eff}+3b_{\rm eff}/x_{\rm FO})},
\end{eqnarray}
where
\begin{eqnarray}
x_{\rm FO} = \ln \frac{0.038 g_{\rm eff} M_{\rm pl} M_1 \langle \sigma_{\rm eff} |v_{\rm rel}| \rangle}{g_\ast^{1/2} x_{\rm FO}^{1/2} }.
\label{Eq:xf}
\end{eqnarray}
%

\begin{figure}[t]
\begin{center}
\includegraphics[clip,width=6.0cm]{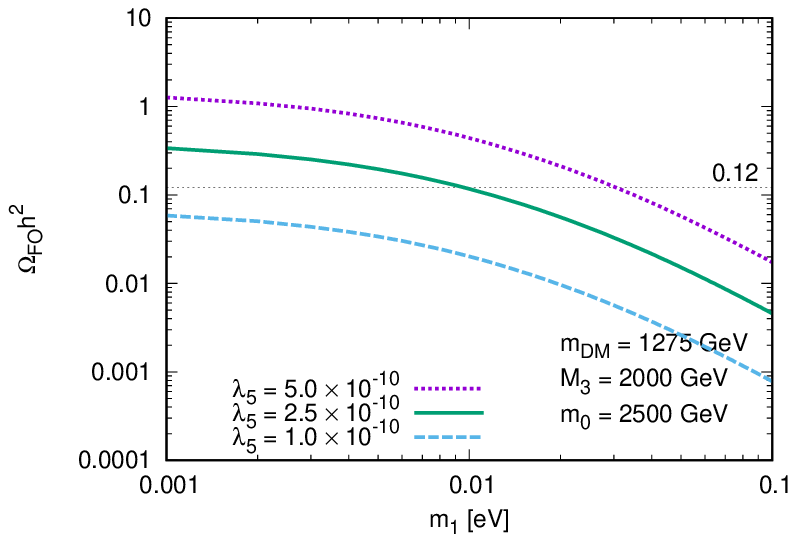}
\includegraphics[clip,width=6.0cm]{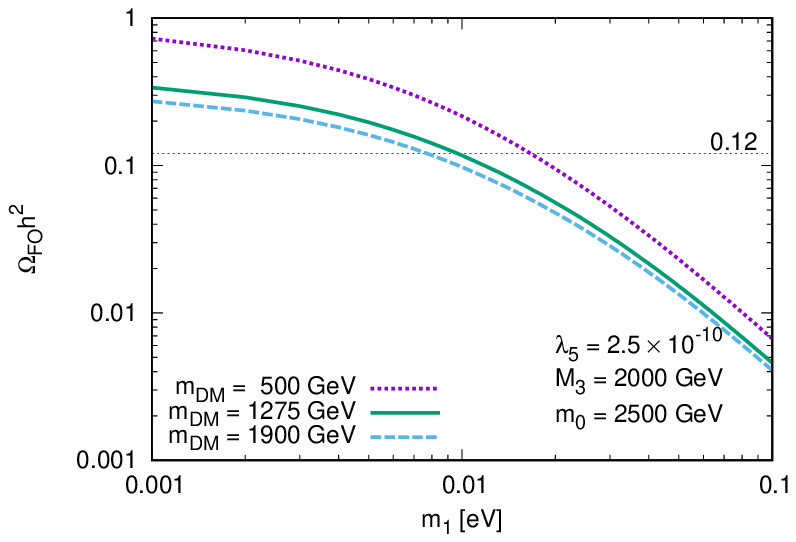} \\
\includegraphics[clip,width=6.0cm]{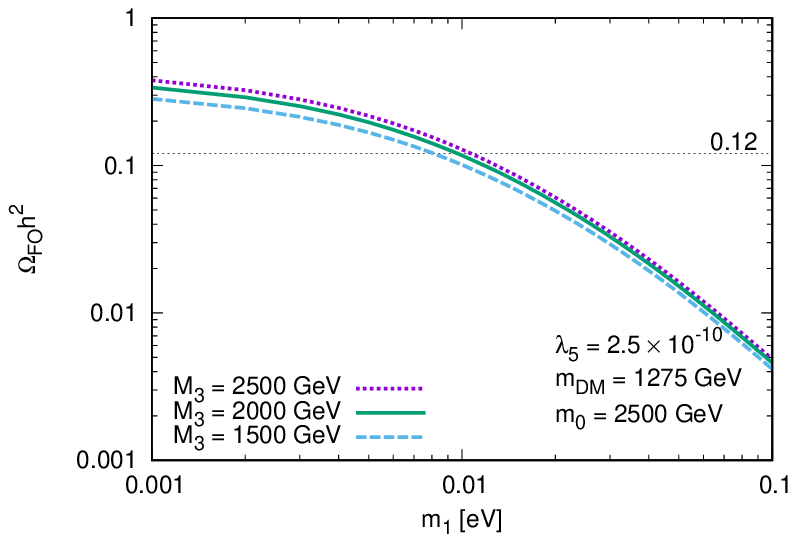} 
\includegraphics[clip,width=6.0cm]{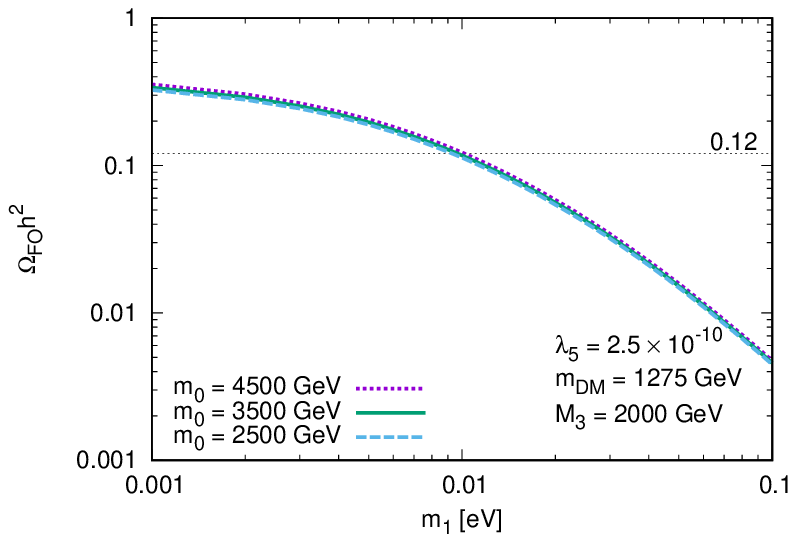}
\caption{The dependence of the predicted relic abundance of dark matter $\Omega_{\rm FO}h^2$ on the lightest neutrino mass $m_1$ in the scotogenic model. The dotted horizontal lines show the observed relic abundance $\Omega_ch^2 = 0.12$. The dependence of $\lambda_5$,  $m_{\rm DM}$,  $M_3$ and $m_0$ on $\Omega_{\rm FO}h^2$ are also shown in the upper left, upper right, lower left and lower right panels, respectively.}
\label{fig:Omegah2_m1_scotogenic}
\end{center}
\end{figure}

\begin{figure}[t]
\begin{center}
\includegraphics[clip,width=6.0cm]{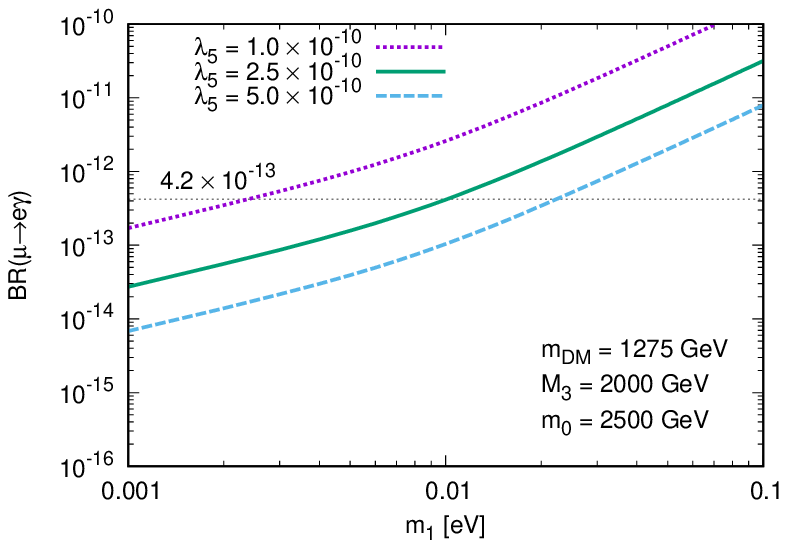}
\includegraphics[clip,width=6.0cm]{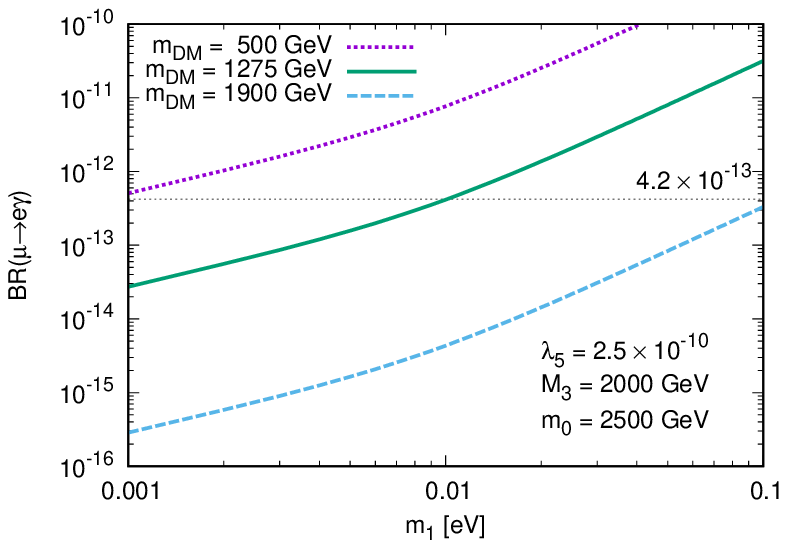}\\
\includegraphics[clip,width=6.0cm]{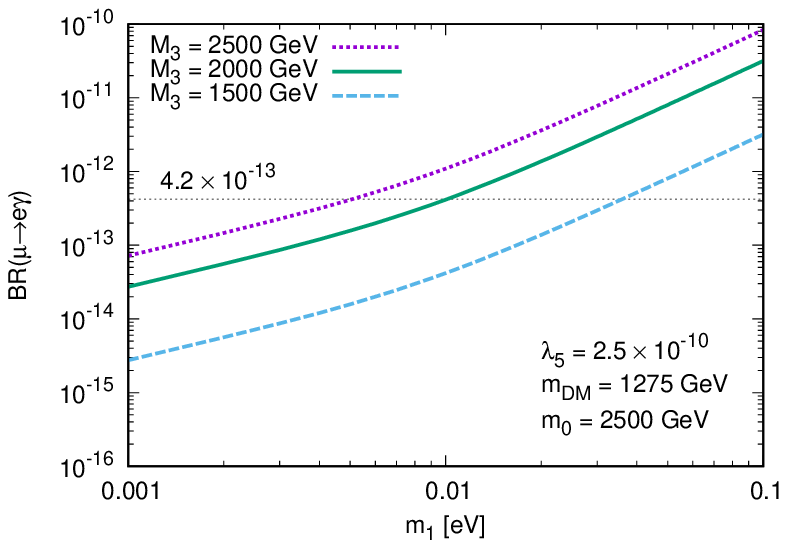}
\includegraphics[clip,width=6.0cm]{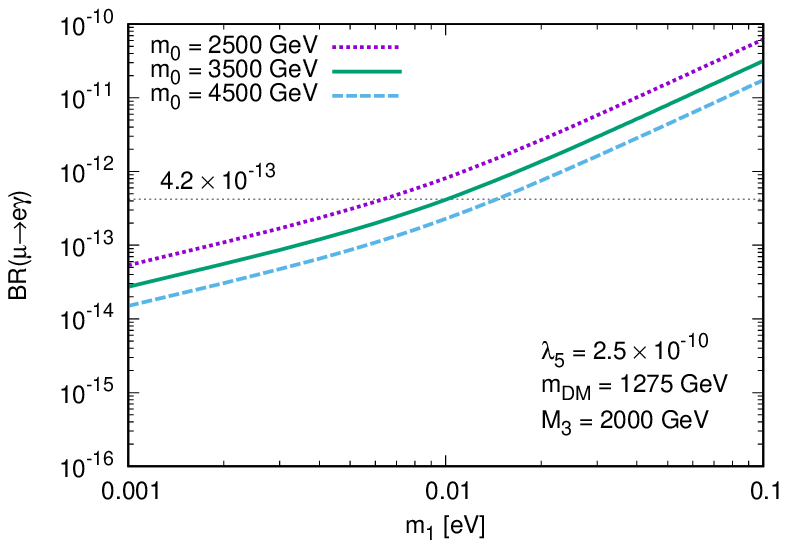}
\caption{The dependence of the predicted branching ratio ${\rm BR}(\mu\rightarrow e\gamma)$ on the lightest neutrino mass $m_1$ in the scotogenic model. The dotted horizontal lines show the observed upper limit of the branching ratio ${\rm BR}(\mu\rightarrow e\gamma) = 4.2 \times 10^{13}$. The dependence of $\lambda_5$,  $m_{\rm DM}$,  $M_3$ and $m_0$ on ${\rm BR}(\mu\rightarrow e\gamma)$ are also shown in the upper left, upper right, lower left and lower right panels, respectively.}
\label{fig:BR_m1_scotogenic}
\end{center}
\end{figure}

\subsection{Parameters in the scotogenic model}
We show the appropriate values of the parameters in the scotogenic model.

For the neutrino sector in the scotogenic model, we can use the data from neutrino oscillation experiments. The neutrino mass ordering (either the normal mass ordering $m_1<m_2<m_3$ or the inverted mass ordering $m_3 < m_1< m_2$) is unsolved problems. The best-fit values of the squared mass differences $\Delta m_{ij}^2=m_i^2-m_j^2$, the mixing angles and the Dirac CP phase for normal mass ordering (with KS atmospheric data) are estimated as \cite{Esteban2020JHEP}
\begin{eqnarray} 
\theta_{12}/^\circ &=& 33.44 \quad (31.27 - 35.86), \quad
\theta_{23}/^\circ = 49.2 \quad (40.1 - 51.7), \nonumber \\
\theta_{13}/^\circ &=& 8.57\quad (8.20 - 8.93), \quad
\delta/^\circ = 197\quad (120 - 369), \nonumber \\
\Delta m^2_{21}/(10^{-5} {\rm eV}^2) &=& 7.42 \quad (6.82- 8.04), \nonumber \\
\Delta m^2_{31}/(10^{-3}{\rm eV}^2) &=& 2.517\quad (2.435 - 2.598),
\label{Eq:neutrino_observation_NO}
\end{eqnarray}
where the parentheses denote the $3 \sigma$ region. On the other hand, in the inverted mass ordering, the squared mass differences and the mixing angles are estimated as \cite{Esteban2020JHEP}
\begin{eqnarray} 
\theta_{12}/^\circ &=& 33.45 \quad (31.27 - 35.87), \quad
\theta_{23}/^\circ = 49.3 \quad (40.3 - 51.8), \nonumber \\
\theta_{13}/^\circ &=& 8.60\quad (8.24 - 8.96), \quad
\delta/^\circ = 282\quad (193 - 352), \nonumber \\
\Delta m^2_{21}/(10^{-5} {\rm eV}^2) &=& 7.42 \quad (6.82- 8.04), \nonumber \\
-\Delta  m^2_{32}/(10^{-3}{\rm eV}^2) &=& 2.498\quad (2.581 - 2.414).
\label{Eq:neutrino_observation_IO}
\end{eqnarray}
Although the neutrino mass ordering is not determined, we assume the normal mass ordering (NO) for the neutrinos. Global analysis recently showed a preference for the normal mass ordering, coming mostly from neutrino oscillation measurements \cite{Salas2018FASS,Salas2018PLB}, however, further updated data has again made the picture less clear \cite{Esteban2020JHEP}.

We take the neutrino masses $m_2, m_3$, mixing angles, Dirac CP phase and Majorana CP phases for our numerical calculations as follows:
\begin{eqnarray}
m_2 &=& \sqrt{m_1^2 + \Delta m^2_{21}|_{\rm best-fit,NO}}, \quad
m_3 = \sqrt{m_1^2 + \Delta m^2_{31}|_{\rm best-fit,NO}}, \nonumber \\
\theta_{ij} &=& \theta_{ij}^{\rm best-fit,NO}, \quad \delta = \delta^{\rm best-fit,NO},\quad
\alpha_i = 0.
\label{Eq:setup_m2_m3_theta_delta_alpha}
\end{eqnarray}
The relic density of the dark matter depends only weakly on CP-violating phases, we neglect the Majorana CP phases \cite{Boer2020PRD}. In this setup, the lightest neutrino mass $m_1$ is still free parameter for the neutrino sector in the scotogenic model. We take
\begin{eqnarray}
m_1= 0.001 - 0.1 \ {\rm eV}.
\label{Eq:setup_m1}
\end{eqnarray}

For the dark sector in the scotogenic model, we adopt the following standard criteria (see, e.g., Refs. \cite{Kubo2006PLB,Ibarra2016PRD,Lindner2016PRD}). 
\begin{itemize}
\item  The quartic coupling satisfies the relation $| \lambda | \ll 1$ for small neutrino masses.
\item Since we assumed that the additional lightest Majorana fermion $N_1$ is the dark matter particles, we require $M_1\lesssim M_2 < M_3$. 
\item The mass scale of new fields is a few TeV. 
\end{itemize}
Hereafter, we use $m_{\rm DM} = M_1 \simeq M_2$. For sake of simplicity and to obtain appropriate branching ratio of $\mu \rightarrow e\gamma$ process, we take 
\begin{eqnarray}
\lambda_5 = 1.0 \times 10^{-11} - 1.0 \times 10^{-9}, \quad
m_{\rm DM}= 500 - 1900  \ {\rm GeV}, 
\label{Eq:setup_lambda5_mDM}
\end{eqnarray}
and
\begin{eqnarray}
M_3 = 1500 - 2500 \ {\rm GeV}, \quad
m_0 = 2500 - 4500 \ {\rm GeV}.
\label{Eq:setup_M3_m0}
\end{eqnarray}
to provide a typical example range.

For illustration, let us consider a benchmark parameter set 
\begin{eqnarray}
m_1 &=& 0.01 \ {\rm eV}, \quad
\lambda_5 = 2.5 \times 10^{-10}, \quad
m_{\rm DM} = 1275 \ {\rm GeV}, \nonumber \\
M_3 &=& 2000 \ {\rm GeV}, \quad
m_0 = 3500 \ {\rm GeV},
\end{eqnarray}
with Eq.(\ref{Eq:setup_m2_m3_theta_delta_alpha}). Using these benchmark values, we obtain $x_{\rm FO} = 24.5$ and 
\begin{eqnarray}
\Omega_{\rm FO} h^2&=& 0.118, \quad
{\rm BR}(\mu\rightarrow e\gamma) = 4.16 \times 10^{-13}, \nonumber \\
 | M_{ee}|  &=& 0.0117,\quad
 \sum m_i= 0.0744,
\end{eqnarray}
which are consistent with observations. The observed dark matter relic abundance is \cite{Planck2020AA}
\begin{eqnarray}
\Omega_{\rm c} h^2 =0.12.
\end{eqnarray}
The measured upper limit of lepton flavor violating $\mu\rightarrow e\gamma$, $\tau\rightarrow \mu\gamma$ and $\tau\rightarrow e\gamma$ processes are \cite{MEG2016EPJC,BABAR2010PRL} 
\begin{eqnarray}
{\rm BR}(\mu\rightarrow e\gamma) &\le& 4.2\times 10^{-13},  \quad
{\rm BR}(\tau\rightarrow \mu\gamma) \le 4.4\times 10^{-8},\nonumber \\ 
{\rm BR}(\tau\rightarrow e\gamma) &\le& 3.3\times 10^{-8}.
\end{eqnarray}
We only account for ${\rm Br}(\mu\rightarrow e\gamma)$ since it is the most stringent constraint. Moreover, we have the following constraints, 
\begin{eqnarray}
| M_{ee}| < 0.066 - 0.155 \ {\rm eV},
\end{eqnarray}
from the neutrinoless double beta decay experiments \cite{GERDA2019Science,Capozzi2020PRD}  and
\begin{eqnarray}
\sum m_i < 0.12 - 0.69 \ {\rm eV},
\end{eqnarray}
from observation of cosmic microwave background radiation \cite{Planck2020AA,Giusarma2016PRD,Vagnozzi2017PRD,Giusarma2018PRD,Capozzi2020PRD}.

The results from a more general parameter search for the scotogenic model are shown in Fig. \ref{fig:Omegah2_m1_scotogenic} and Fig. \ref{fig:BR_m1_scotogenic}.

Figure \ref{fig:Omegah2_m1_scotogenic} shows the dependence of the predicted relic abundance of dark matter $\Omega_{\rm FO}h^2$ on the lightest neutrino mass $m_1$ in the scotogenic model. The dotted horizontal lines show the observed relic abundance $\Omega_ch^2 = 0.12$. Figure \ref{fig:BR_m1_scotogenic} shows the dependence of the predicted branching ratio ${\rm BR}(\mu\rightarrow e\gamma)$ on the lightest neutrino mass $m_1$ in the scotogenic model. The dotted horizontal lines show the observed upper limit of the branching ratio ${\rm BR}(\mu\rightarrow e\gamma) = 4.2 \times 10^{13}$. The dependence of $\lambda_5$,  $m_{\rm DM}$,  $M_3$ and $m_0$ on $\Omega_{\rm FO}h^2$ in Fig.\ref{fig:Omegah2_m1_scotogenic} and on ${\rm BR}(\mu\rightarrow e\gamma)$ in Fig. \ref{fig:BR_m1_scotogenic} are also shown in the upper left, upper right, lower left and lower right panels, respectively. Fig. \ref{fig:Omegah2_m1_scotogenic} and Fig. \ref{fig:BR_m1_scotogenic} show that our choice of parameters in Eqs. (\ref{Eq:setup_m2_m3_theta_delta_alpha}), (\ref{Eq:setup_m1}), (\ref{Eq:setup_lambda5_mDM}) and (\ref{Eq:setup_M3_m0}) is appropriate for the scotogenic model.

Hereafter, we set 
\begin{eqnarray}
M_3 = 2000 \ {\rm GeV}, \quad
m_0 = 3500 \ {\rm GeV},
\label{Eq:setup_M3_m0_typical}
\end{eqnarray}
as a typical values in this section.

\section{PBHs and Scotogenic dark matter \label{sec:PBHsAndScotogenic}}

\subsection{Freeze-out effect on PBH evaporation}
\begin{figure}[t]
\begin{center}
\includegraphics{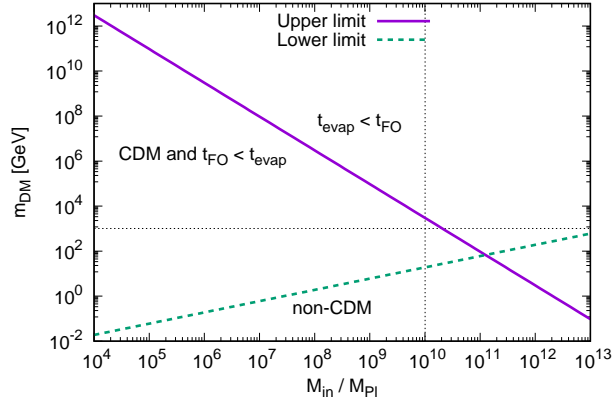}
\caption{The upper and lower limit of the mass of dark matter $m_{\rm DM}$ for the initial mass of PBHs $M_{\rm in}/M_{\rm Pl}$. The dotted horizontal line shows the typical mass of the dark matter in our setup in this section ($m_{\rm DM} = 1000$ GeV). The dotted vertical line shows the order of the upper limits of $M_{\rm in}/M_{\rm Pl}$ for $m_{\rm DM} = 1000$ GeV.}
\label{fig:PBH_mDM_Upper_Lower_MinPMpl}
\end{center}
\end{figure}

According to Gondolo et al. \cite{Gondolo2020PRD}, if the PBH evaporation ends before the freeze-out of the dark matter, then the dark matter particles produced by Hawking radiation may reach chemical equilibrium with the bath particles, due to the enormous number of thermally produced dark matter particles, and dark matter particles from PBHs evaporation give no extra contribution to the relic abundance of the dark matter. On the other hand, if PBH evaporate after the freeze-out of the dark matter, then the dark matter particles produced by Hawking radiation may neither thermalize with the bath particles nor annihilate with each other. Thus, if $t_{\rm FO} < t_{\rm evap}$, then the dark matter particles produced from PBHs may contribute to the final relic abundance of the dark matter. The effect of PBH evaporation on dark matter relic abundance can be summarized as follows: 
\begin{eqnarray}
\begin{cases}
\Omega_{\rm FO}h^2 \le \Omega_{\rm c}h^2 & {\rm for} \  t_{\rm evap} < t_{\rm FO} \\
\Omega_{\rm FO}h^2 + \Omega_{\rm PBH}h^2 \le \Omega_{\rm c}h^2 &  {\rm for} \  t_{\rm FO} < t_{\rm evap}
\end{cases}.
\end{eqnarray}

For our purpose, we concentrate our discussion for the case of $t_{\rm FO} < t_{\rm evap}$. The requirement of $t_{\rm FO} < t_{\rm evap}$ is equivalent to the requirement of $T_{\rm FO} > T_{\rm evap}$. This requirement translates into the upper limit of the dark matter mass 
\begin{eqnarray}
\left(\frac{m_{\rm DM}} {\rm GeV}\right) \lesssim 3.00 \times 10^{18} \left( \frac{x_{\rm FO}}{25}\right) \left(\frac{g_* (T_{\rm BH})}{106.75}\right)^{1/4}   \left( \frac{M_{\rm Pl}}{M_{\rm in}}\right)^{3/2},
\label{Eq:mDMUpperFromTfoTevap}
\end{eqnarray}
by Eq. (\ref{Eq:Tevap}) and Eq. (\ref{Eq:xf25}) [Eq. (\ref{Eq:xf})].

\subsection{$\beta < \beta_{\rm c}$ case \label{section:PBHsAndScotogenic_RD}}
Now, we include the effect of the scotogenic dark matter on the PBHs into our discussion in the case of $\beta < \beta_{\rm c}$.

Figure \ref{fig:PBH_mDM_Upper_Lower_MinPMpl} shows the upper and lower limit of the mass of dark matter $m_{\rm DM}$ for the initial mass of PBHs $M_{\rm in}/M_{\rm Pl}$. For example. some specific values of $M_{\rm in}/M_{\rm Pl}$ yield
\begin{eqnarray}
\left(\frac{m_{\rm DM}} {\rm GeV}\right)  
 \simeq
\begin{cases}
 0.019 -  3.0 \times 10^{12} & \ {\rm for} \  \frac{M_{\rm in}}{M_{\rm Pl}} = 1\times 10^4 \\
 19 -  3000 & \ {\rm for} \   \frac{M_{\rm in}}{M_{\rm Pl}} = 1\times 10^{10} \\
 60.1 - 94.9 & \ {\rm for} \ \frac{M_{\rm in}}{M_{\rm Pl}} = 1\times 10^{11} \\
 67 & \ {\rm for} \ \frac{M_{\rm in}}{M_{\rm Pl}} = 1.257\times 10^{11} \\
\end{cases},\nonumber 
\end{eqnarray}
where $m_{\rm DM} \simeq 67$ GeV for $M_{\rm in}/M_{\rm Pl} = 1.257\times 10^{11}$ is obtained at the cross point of the upper limit and the lower limit as shown in Fig. \ref{fig:PBH_mDM_Upper_Lower_MinPMpl}. Thus, we have
\begin{eqnarray}
10^4 \lesssim \frac{M_{\rm in}}{M_{\rm Pl}} \lesssim 1.257 \times 10^{11} \left( \frac{x_{\rm FO}}{25}\right)^{1/2} \left(\frac{g_* (T_{\rm BH})}{106.75}\right)^{1/8}.
\label{Eq:MinPMplGeneral}
\end{eqnarray}
The upper limit of the $M_{\rm in}/M_{\rm Pl}$ in Eq. (\ref{Eq:MinPMplGeneral}) is obtained by R.H.S of Eq. (\ref{Eq:mDMlowerFromWDM}) = R.H.S of Eq. (\ref{Eq:mDMUpperFromTfoTevap}). We recall that in the warm-plus-cold dark case, there is a slow dark matter component and the WDM constraint will be weaker \cite{Baldes2020JCAP,Boyarsky2009JCAP,Baur2017JCAP}. We note that Eq. (\ref{Eq:MinPMplGeneral}) is general, model independent, constraint of the initial mass of PBH for $t_{\rm FO} < t_{\rm evap}$.

We would like to comment that the WDM constraint assumes no elastic scatterings. Even if $N_1N_1\leftrightarrow \bar{f}f$ interactions are negligible, elastic scatterings $Nf \leftrightarrow Nf$ mediated by t-channel inert doublet exchange, may still be allowed. See Refs. \cite{Arcadi2011PRD,Aarssen2012PRD,Bringmann2016PRD,Cirelli2017JCAP,Binder2017PRD,Duch2017JCAP} for example of the kinetic decoupling in dark matter models. From the discussion of the direct detection of fermion dark matter \cite{Ibarra2016PRD}, of exploring X-ray lines \cite{Faisel2014PLB} and of collider search experiments \cite{Ho2013PRD,Ho2014PRD,Ahriche2020PRD} in the scotogenic model, we can expect that the size of such effects is small enough to ascertain in which areas of parameter space the WDM constraint can be applied in the scotogenic model.

The dotted horizontal line in Fig. \ref{fig:PBH_mDM_Upper_Lower_MinPMpl} shows the typical mass of the dark matter in our setup in this section ($m_{\rm DM} = 1000$ GeV). The dotted vertical line shows the order of the upper limits of $M_{\rm in}/M_{\rm Pl}$ as
\begin{eqnarray}
\left( \frac{M_{\rm in}}{M_{\rm Pl}} \right)_{\rm max} \simeq 10^{10} & \ {\rm for} \ m_{\rm DM} = 1000 \ {\rm GeV}
\label{fig:PBH_Scoto_Upper_MinPMpl}
\end{eqnarray}
According to Eq. (\ref{Eq:MinPMplGeneral}) and Eq. (\ref{fig:PBH_Scoto_Upper_MinPMpl}), we conclude conservatively that the initial PBHs mass should be
\begin{eqnarray}
10^4 \lesssim \frac{M_{\rm in}}{M_{\rm Pl}} \lesssim 10^{10} \left( \frac{x_{\rm FO}}{25}\right)^{1/2} \left(\frac{g_* (T_{\rm BH})}{106.75}\right)^{1/8},
\label{Eq:MinPMplConstraint}
\end{eqnarray}
with the $\mathcal{O}$(1) TeV scogtogenic dark matter. For $x_{\rm FO} = 25$ and $g_*(T_{\rm BH})=106.75$, we have
\begin{eqnarray}
10^4 \lesssim \frac{M_{\rm in}}{M_{\rm Pl}} \lesssim 10^{10}.
\label{Eq:MinPMplConstraint_2}
\end{eqnarray}
This is the first main result in this paper.

The relic abundance of the dark matter produced from PBHs in radiation dominated era, $\beta < \beta_{\rm c}$, is estimated by Eqs. (\ref{Eq:OmegaPBHh2_RD_T>m}) for $T_{\rm BH}^{\rm in} > m_{\rm DM}$ or (\ref{Eq:OmegaPBHh2_RD_T<m}) for $T_{\rm BH}^{\rm in} < m_{\rm DM}$. On the other hand, in the  matter dominated (PBH dominated) era, $\beta > \beta_{\rm c}$,  Eqs. (\ref{Eq:OmegaPBHh2_MD_T>m}) for $T_{\rm BH}^{\rm in} > m_{\rm DM}$ or (\ref{Eq:OmegaPBHh2_MD_T<m}) $T_{\rm BH}^{\rm in} < m_{\rm DM}$ should be used to estimate the relic abundance of dark matter. Before calculations of the relic abundance of the scotogenic dark matter from PBHs, we should evaluate $T_{\rm BH}^{\rm in}$ and $\beta$ for $m_{\rm DM} = \mathcal{O}(1)$ TeV.

\begin{figure}[t]
\begin{center}
\includegraphics{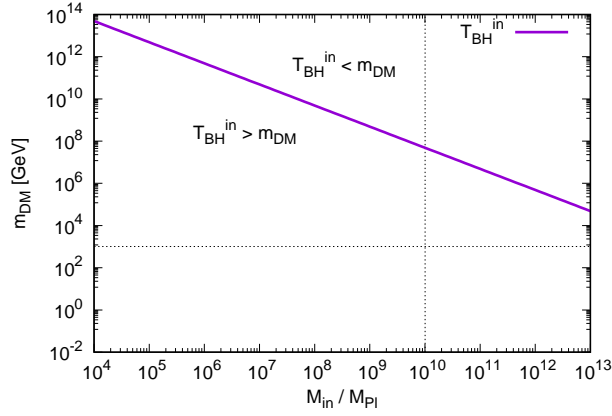}
\caption{The region for $T_{\rm BH}^{\rm in} > m_{\rm DM}$ and for $T_{\rm BH}^{\rm in} < m_{\rm DM}$ in the $m_{\rm DM}$ - $M_{\rm in}/M_{\rm Pl}$ space for $\beta < \beta_{\rm c}$ case. The dotted horizontal line shows the typical mass of the dark matter in our setup in this section ($m_{\rm DM} = 1000$ GeV). The dotted vertical line shows the order of the upper limits of $M_{\rm in}/M_{\rm Pl}$ for $m_{\rm DM} = 1000$ GeV. The relation $T_{\rm BH}^{\rm in} > m_{\rm DM}$ should be satisfied with $m_{\rm DM} = 1000$ GeV. 
}
\label{fig:PBH_Tin_MinPMpl}
\end{center}
\end{figure}

Figure \ref{fig:PBH_Tin_MinPMpl} shows the region for $T_{\rm BH}^{\rm in} > m_{\rm DM}$ and for $T_{\rm BH}^{\rm in} < m_{\rm DM}$ in the $m_{\rm DM}$ - $M_{\rm in}/M_{\rm Pl}$ space for $\beta < \beta_{\rm c}$ case where $T_{\rm BH}^{\rm in}$ is estimated by Eq. (\ref{Eq:PBH_Tin_MinPMpl}). The dotted horizontal line shows the typical mass of the dark matter in our setup in this section ($m_{\rm DM} = 1000$ GeV). The dotted vertical line shows the order of the upper limits of $M_{\rm in}/M_{\rm Pl}$ for $m_{\rm DM} = 1000$ GeV. We see that the relation $T_{\rm BH}^{\rm in} > m_{\rm DM}$ is satisfied with $m_{\rm DM} = 1000$ GeV.

\begin{figure}[t]
\begin{center}
\includegraphics{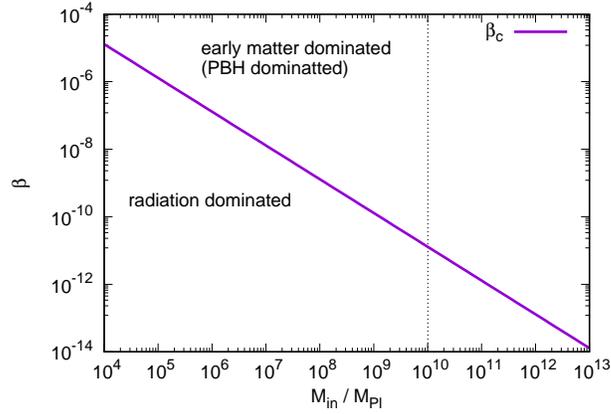}
\caption{The region for the radiation dominated era, $\beta < \beta_{\rm c}$ and for the matter (PBH) dominated era, $\beta > \beta_{\rm c}$, in the $\beta$ - $M_{\rm in}/M_{\rm Pl}$. The dotted vertical line shows the order of the upper limits of $M_{\rm in}/M_{\rm Pl}$ for $m_{\rm DM} = 1000$ GeV.}
\label{fig:PBH_betac_MinPMpl}
\end{center}
\end{figure}

Figure \ref{fig:PBH_betac_MinPMpl} shows the region for the radiation dominated era, $\beta < \beta_{\rm c}$ and for the matter (PBH) dominated era, $\beta > \beta_{\rm c}$, in the $\beta$ - $M_{\rm in}/M_{\rm Pl}$ space where $\beta_{\rm c}$ is estimated by Eq. (\ref{Eq:RDconstraint}) for $\gamma = 0.2$ and $g_*(T_{\rm BH}) = 106.75$. As same as Figs. \ref{fig:PBH_mDM_Upper_Lower_MinPMpl} and \ref{fig:PBH_Tin_MinPMpl}, the dotted vertical line shows the order of the upper limits of $M_{\rm in}/M_{\rm Pl}$ for $m_{\rm DM} = 1000$ GeV. Figure \ref{fig:PBH_betac_MinPMpl} shows that we can think the PBHs evaporated in the radiation dominated era in the early Universe as an appropriate scenario for scotogenic DM to be sourced from a mix of coannihilations and PBH decays. In this case, the relic abundance of scotogenic dark matter from PBHs can be estimated by Eq. (\ref{Eq:OmegaPBHh2_RD_T>m}).

For illustration, let us consider a benchmark parameter set
\begin{eqnarray}
m_1 &=& 0.03 \ {\rm eV}, \quad
\lambda_5 = 2.5 \times 10^{-10}, \quad
m_{\rm DM} = 1670 \ {\rm GeV}, \nonumber \\
M_3 &=& 2000 \ {\rm GeV}, \quad
m_0 = 3500 \ {\rm GeV},
\label{Eq:setup_M1=0.03_mDM_M3_m0}
\end{eqnarray}
for the scotogenic model with Eq.(\ref{Eq:setup_m2_m3_theta_delta_alpha}) and 
\begin{eqnarray}
\gamma &=& 0.2, \quad
g_*(T_{\rm in}) =  g_*(T_{\rm BH})=106.75 + \frac{7}{8} g_{\rm DM},  \quad
g_{\rm DM} =  2,  \quad
C_{\rm DM} =  3/4,
\label{Eq:setup_gamma_g*Tin_gDM_CDM}
\end{eqnarray}
as well as
\begin{eqnarray}
\beta = 5.4 \times 10^{-16},  \quad
M_{\rm in}/M_{\rm Pl} = 1 \times 10^{10},
\end{eqnarray}
for the PBHs. Using these benchmark values, we obtain $x_{\rm FO} = 26.15 $ and 
\begin{eqnarray}
\Omega_{\rm FO}h^2 &=& 0.0295, \quad
\Omega_{\rm PBH}h^2 = 0.0907, \quad
\Omega_{\rm DM}h^2 = \Omega_{\rm FO}h^2  + \Omega_{\rm PBH}h^2 = 0.12, \nonumber \\
{\rm BR}(\mu\rightarrow e\gamma) &=& 4.08\times 10^{-13}, \quad
 | M_{ee}|  = 0.031, \quad
 \sum m_i= 0.12, 
\end{eqnarray}
which are consistent with observations.

\begin{figure}[t]
\begin{center}
\includegraphics{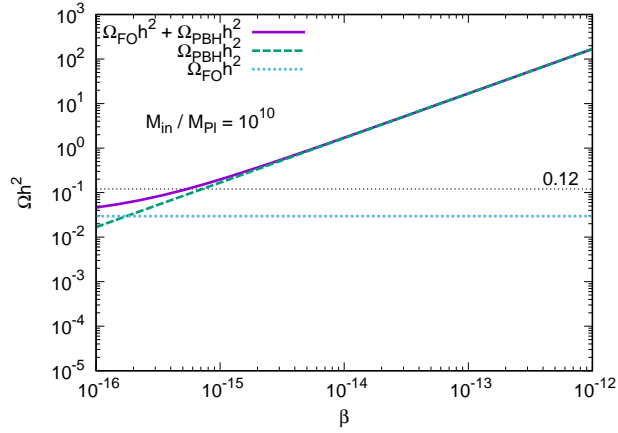}
\caption{The dependence of the predicted relic abundance of dark matter $\Omega h^2$ on the initial density of PBHs $\beta$ for $\beta < \beta_{\rm c}$ case. The dotted horizontal line shows the observed relic abundance $\Omega_ch^2 = 0.12$. We take $M_{\rm in}/M_{\rm Pl}=10^{10}$. }
\label{fig:scoto_PBH_Omegah2_beta_MinPMpl}
\end{center}
\end{figure}

\begin{figure}[t]
\begin{center}
\includegraphics{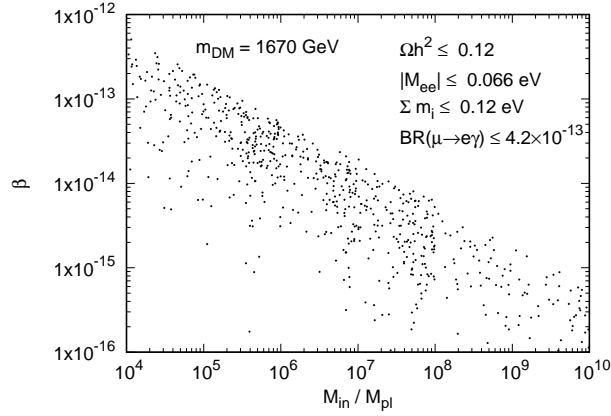}
\caption{Allowed initial density of PBHs $\beta$ and initial mass of PBHs $M_{\rm in}/M_{\rm Pl}$.}
\label{fig:scoto_PBH_allowed_omegah2_MinPMpl_M1}
\end{center}
\end{figure}

The results from a more general parameter search for the scotogenic model and PBH are shown in Fig. \ref{fig:scoto_PBH_Omegah2_beta_MinPMpl} and Fig. \ref{fig:scoto_PBH_allowed_omegah2_MinPMpl_M1}.
 
Figure \ref{fig:scoto_PBH_Omegah2_beta_MinPMpl} shows the dependence of the predicted relic abundance of dark matter $\Omega h^2$ on the initial density of PBHs $\beta$ for $\beta < \beta_{\rm c}$ case where we take the values in Eqs. (\ref{Eq:setup_m2_m3_theta_delta_alpha}), (\ref{Eq:setup_M1=0.03_mDM_M3_m0}) and (\ref{Eq:setup_gamma_g*Tin_gDM_CDM}). The dotted horizontal line shows the observed relic abundance $\Omega_{\rm c}h^2 = 0.12$. We take $M_{\rm in}/M_{\rm Pl}=10^{10}$.  

Recall that, without scotogenic dark matter, the upper limit of the $\beta$ is obtained as $\beta  \lesssim 0.129  (M_{\rm in}/M_{\rm Pl})^{-1}$ [Eq.(\ref{Eq:RDconstraint})] as the RD constraint and we obtain 
\begin{eqnarray}
\beta  \lesssim 
10^{-11} & \ {\rm for} \ M_{\rm in}/M_{\rm Pl}=10^{10}.
\end{eqnarray}
On the other hand, if we include the effect of the scotogenic dark matter on PBHs into our discussion, we obtain 
\begin{eqnarray}
\beta  \lesssim 
10^{-15} & \ {\rm for} \ M_{\rm in}/M_{\rm Pl}=10^{10},
\end{eqnarray}
as shown in Fig. \ref{fig:scoto_PBH_Omegah2_beta_MinPMpl}. We can understand this behaviour of $\beta$ in Fig. \ref{fig:scoto_PBH_Omegah2_beta_MinPMpl} by  Eq. (\ref{Eq:OmegaPBHh2_RD_T>m}). Since $\Omega_{\rm PBH}  h^2 \propto \beta ( M_{\rm in}/M_{\rm Pl} )^{1/2}$, we have a relation $\beta \propto ( M_{\rm in}/M_{\rm Pl} )^{-1/2}$. Moreover, due to the existence of the scotogenic dark matter from freeze-out, $\Omega_{\rm PBH}  h^2$ should be decrease with $\Omega_{\rm FO}  h^2$. Thus, the upper limit of the initial PBHs density $\beta$ should decrease for coexistence the scotogenic dark matter from freeze-out and the scotogenic dark matter from PBHs.

Figure  \ref{fig:scoto_PBH_allowed_omegah2_MinPMpl_M1} shows the allowed initial density of PBHs $\beta$ and initial mass of PBHs $M_{\rm in}/M_{\rm Pl}$ for 
\begin{eqnarray}
&& \Omega h^2 = \Omega_{\rm FO} h^2 + \Omega_{\rm PBH} h^2  \le 0.12, \quad
 |M_{ee}| \le 0.066\ {\rm eV}, \nonumber \\
&& \sum m_i \le 0.12\ {\rm eV}, \quad
 {\rm BR}(\mu\rightarrow e\gamma)  \le 4.2 \times 10^{-13},
\end{eqnarray}
and
\begin{eqnarray}
&& m_1 = 0.001 - 0.1 \ {\rm eV}, \quad
\lambda_5 = 1.0 \times 10^{-11} - 1.0 \times 10^{-9}, \nonumber \\
&& m_{\rm DM} = 1675 \ {\rm GeV} \quad
 M_3 = 2000 \ {\rm GeV}, \quad
 m_0 = 3500 \ {\rm GeV},
\end{eqnarray}
with Eqs. (\ref{Eq:setup_m2_m3_theta_delta_alpha}) and (\ref{Eq:setup_gamma_g*Tin_gDM_CDM}). We comment that the upper limit of the initial PBHs density $\beta$ depends only weakly on the mass of dark matter $m_{\rm DM}$ in the realistic scotogenic model as shown in Fig. \ref{fig:scoto_PBH_allowed_omegah2_MinPMpl_M1}. Although we have $\beta \propto m_{\rm DM}^{-1}( M_{\rm in}/M_{\rm Pl} )^{-1/2}$ by Eq. (\ref{Eq:OmegaPBHh2_RD_T>m}), the realistic scotogenic model should keep $m_{\rm DM} = $ a few TeV and variation of $m_{\rm DM}^{-1}$ should be small. For example, we obtain $m_{\rm DM}^{-1}$ = 0.00069 - 0.00052 GeV$^{-1}$ for $m_{\rm DM}$ = 1450  - 1900 GeV. From Fig. \ref{fig:scoto_PBH_allowed_omegah2_MinPMpl_M1}, we obtain the following upper limit of the initial PBHs density 
\begin{eqnarray}
\beta \lesssim 5 \times 10^{-11} \left( \frac{M_{\rm in}}{M_{\rm Pl}} \right)^{-1/2}.
\end{eqnarray}
for $\mathcal{O}$(1) TeV scotogenic dark matter. This is the second main result in this paper.

\subsection{$\beta > \beta_{\rm c}$ case}
In the previous subsection, we assume that the PBHs have evaporated in radiation dominated era ($\beta < \beta_{\rm c}$). In this subsection, we assume that the PBH evaporation happens in matter dominated era, PBH dominated era ($\beta > \beta_{\rm c}$).

If $\beta > \beta_c$, the entropy produced by the PBH decay may allow more freedom for the dark matter parameter space in the scotogenic model. This entropy production is a significant factor \cite{Fujita2014PRD, Masina2020EPJP, Baldes2020JCAP,Chaudhuri202011arXiv}, which may dilute any dark matter produced thermally in the bath at higher temperatures. Thus, the requirement of $m_{\rm DM} \sim \mathcal{O}$(TeV) for scotogenic dark mater in the usual freeze-out scenario via dark matter (co)-annihilation may not necessary with the addition of the PBHs in the case of $\beta > \beta_c$. To be more precise, we should study the  possibility of acceptance of more light ($m_{\rm DM} \lesssim$ MeV) or more heavy ($m_{\rm DM} \gtrsim$ PeV) scotogenic dark matter with PBHs.

\subsubsection{$m_{\rm DM} \lesssim$ MeV case}
First, we  study the possibility of acceptance of more light scotogenic dark matter ($m_{\rm DM} \lesssim$ MeV). According to Ref. \cite{Fujita2014PRD}, the entropy production dilutes dark matter emitted from PBHs and relaxes the restriction from $\Omega_{\rm PBH}h^2$. Moreover, the entropy production changes the evolution of the scale factor and hence alters the constraint on the warm dark matter. As a result, some lighter dark matter mass regions, which are originally excluded, may be allowed with the entropy production. This is a good news for cogenesis scenario (producing baryon number and dark matter of the same order); however, the light scotogenic dark matter, $m_{\rm DM} < \mathcal{O}$ (TeV), yields unacceptable large branching ration of a lepton number violating process, ${\rm BR}(\mu \rightarrow e\gamma)$, for the appropriate range of lightest neutrino mass $m_1 = 0.001 - 0.1$ eV, as shown in  the upper right panel in Fig. \ref{fig:BR_m1_scotogenic}. The branching ratio ${\rm BR}(\mu \rightarrow e\gamma)$ should be determined without dark matter relic density. Thus, more light scotogenic dark matter ($m_{\rm DM} \lesssim$ MeV) should be excluded if there is an entropy boost due to the PBH decay. 

We would like to note that there is another possibility for realize more lighter scotogenic dark matter. For the case of  $T_{\rm BH}^{\rm in} > m_{\rm DM}$, the light dark matter which conflicts with the WDM constraint if dark matter only has gravitational interactions. However, it may be possible to find parameter space in the scotogenic model, which allows for sufficient elastic scattering following the PBH evaporation, to retard the dark matter velocity and come back into agreement with structure formation (primordial velocity $v_{DM} \lesssim 1.8\times 10^{-8}$)\cite{Arcadi2011PRD,Aarssen2012PRD,Bringmann2016PRD,Cirelli2017JCAP,Binder2017PRD,Duch2017JCAP} (if the dark matter is light, chemical equilibrium should not be re-established which would lead to overproduction, this can be achieved if the inert doublet mass $m_0$ is sufficiently large, and the Yukawa couplings of dark matter are sufficiently small). If sufficient elastic scattering occurs, more lighter scotogenic dark matter may be allowed; however, as we mentioned in \ref{section:PBHsAndScotogenic_RD}, the effects of these elastic scatterings $Nf \leftrightarrow Nf$ mediated by t-channel inert doublet exchange may be small. Even if the effect of these elastic scattering is large,  more lighter scotogenic dark matter may conflict the observed upper limit of the branching ratio ${\rm BR}(\mu \rightarrow e\gamma)$.

We conclude that more light scotogenic dark matter ($m_{\rm DM} \lesssim$ MeV) should be excluded even if there is an additional contribution from the PBH decay. This conclusion is the third advantage in this paper.

\begin{figure}[t]
\begin{center}
\includegraphics{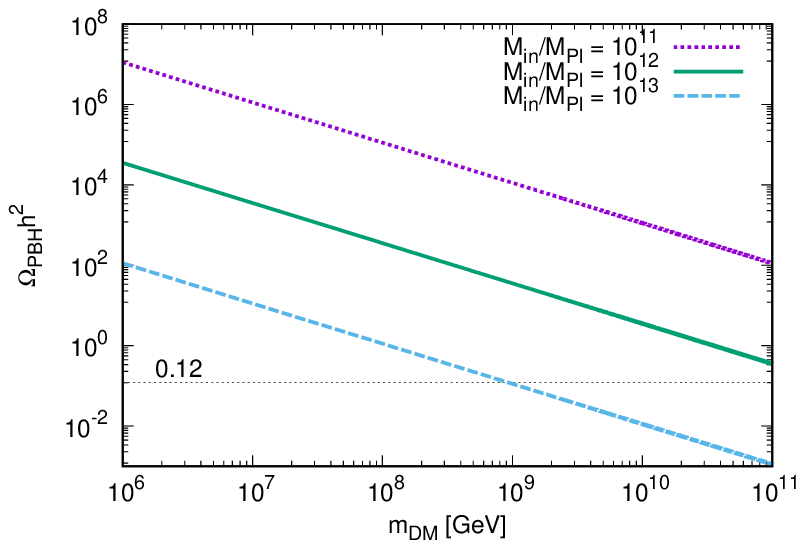}
\includegraphics{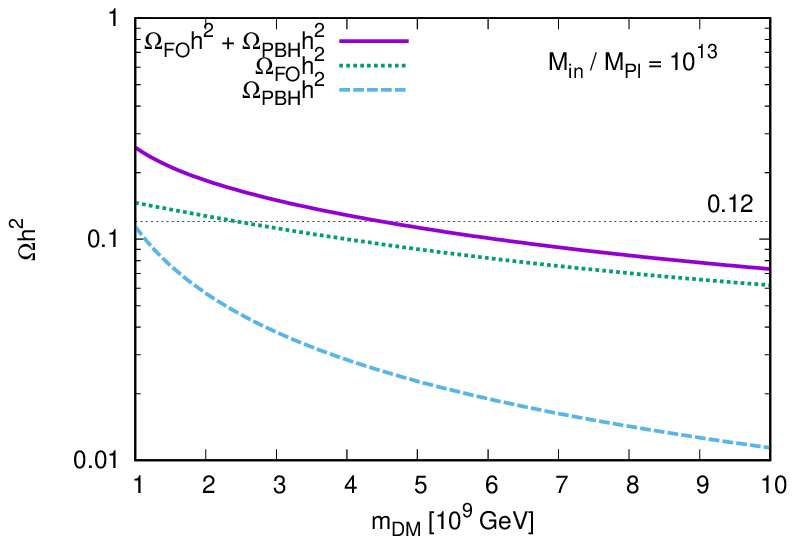}
\caption{The dependence of the predicted relic abundance of dark matter on the mass of scotogenic dark matter in the case of $m_{\rm DM} \gtrsim$ PeV and $\beta > \beta_{\rm c}$. The upper panel shows the predicted relic abundance of dark matter from PBH evaporation. The lower panel shows the total predicted relic abundance of dark matter for $m_{\rm DM} \sim 10^9$ GeV and $M_{\rm in}/M_{\rm Pl} \sim 10^{13}$. The predicted total relic abundance for $m_{\rm DM} \gtrsim 10^9$ GeV and $M_{\rm in}/M_{\rm Pl} \sim 10^{13}$ is consistent with the observed relic density $\Omega_{\rm c} h^2 = 0.12$.}
\label{fig:fig_Omegah2_PBH_DOM}
\end{center}
\end{figure}

\subsubsection{$m_{\rm DM} \gtrsim$ PeV case}
Next, we  study the possibility of acceptance of more heavy scotogenic dark matter ($m_{\rm DM} \gtrsim$ PeV). From Eq. (\ref{Eq:OmegaPBHh2_MD_T<m}), in the case of $T^{\rm in}_{\rm BH} < m_{\rm DM}$, the observed relic density can be achieved with $m_{\rm DM} < M_{\rm Pl}$ provided $M_{\rm in} \gtrsim 10^{10}M_{\rm Pl}$. For $M_{\rm in}/M_{\rm Pl} \sim 10^{13}$ (the BBN bound), we have $m_{\rm DM} \sim 10^9$ GeV. We will show that $m_{\rm DM} \sim 10^9$ GeV may not be too heavy for neutrino mass generation as well as dark matter production in the $\beta > \beta_{\rm c}$ case.

Figure. \ref{fig:fig_Omegah2_PBH_DOM} shows the dependence of the predicted relic abundance of dark matter on the mass of scotogenic dark matter in the case of $m_{\rm DM} \gtrsim$ PeV and $\beta > \beta_{\rm c}$. The dotted horizontal lines show the observed relic abundance $\Omega_{\rm c}h^2 = 0.12$. For illustration, we consider a benchmark parameter set
\begin{eqnarray}
m_1 &=& 0.03 \ {\rm eV}, \quad
\lambda_5 = 2.5 \times 10^{-10}, \nonumber \\
M_3 &=& 2.0 \times 10^{10} \ {\rm GeV}, \quad
m_0 = 3.0 \times 10^{10} \ {\rm GeV},
\end{eqnarray}
for the scotogenic model with Eqs. (\ref{Eq:setup_m2_m3_theta_delta_alpha}) and (\ref{Eq:setup_gamma_g*Tin_gDM_CDM}) in Fig. \ref{fig:fig_Omegah2_PBH_DOM}. We will show the result from more a general parameter search later. 
The upper panel in Fig. \ref{fig:fig_Omegah2_PBH_DOM} shows the dependence of the predicted relic abundance of dark matter from PBH evaporation $\Omega_{\rm PBH} h^2$ on the mass of dark matter $m_{\rm DM}$.  The lower panel in Fig. \ref{fig:fig_Omegah2_PBH_DOM} shows the total predicted relic abundance of dark matter for $m_{\rm DM} \sim 10^9$ GeV and $M_{\rm in}/M_{\rm Pl} \sim 10^{13}$. As we mentioned, the predicted total relic abundance for $m_{\rm DM} \gtrsim 10^9$ GeV and $M_{\rm in}/M_{\rm Pl} \sim 10^{13}$ is consistent with the observed relic density $\Omega_{\rm c} h^2 = 0.12$.

We would like to note that if the freeze-out mechanism is only allowed for dark matter generation, the heavy scotogenic dark matter $m_{\rm DM} \gtrsim 10^9$ GeV could not be main candidate of the dark matter. This situation can be changed if the socotogenic dark matter can be produces by PBH evaporation. The emitted scotogenic dark matters by PBH evaporation compensate for the lack of relic abundance by freeze-out mechanism and the total relic abundance $\Omega_{\rm FO}h^2 + \Omega_{\rm PBH}h^2$ can achieve $0.12$ (the observed value).

\begin{figure}[t]
\begin{center}
\includegraphics{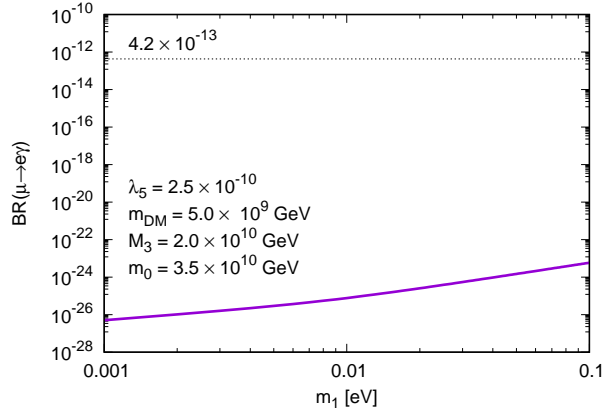}
\caption{The dependence of the predicted branching ratio ${\rm BR}(\mu \rightarrow e\gamma)$ on the lightest neutrino mass $m_1$. The heavy scotogenic dark matter $m_{\rm DM} \gtrsim 10^9$ GeV is consistent with the observed upper limit of the the branching ratio, ${\rm BR}(\mu \rightarrow e\gamma) \le 4.2 \times 10^{-13}$; however, a lepton flavor violating process $\mu \rightarrow  e \gamma$ should be almost completely suppressed. }
\label{fig:Scoto_BR_m1_PBHDOM}
\end{center}
\end{figure}

Figure. \ref{fig:Scoto_BR_m1_PBHDOM} shows that the dependence of the predicted branching ratio ${\rm BR}(\mu \rightarrow e\gamma)$ on the lightest neutrino mass $m_1$. The heavy scotogenic dark matter $m_{\rm DM} \gtrsim 10^9$ GeV is consistent with the observed upper limit of the the branching ratio, ${\rm BR}(\mu \rightarrow e\gamma) \le 4.2 \times 10^{-13}$ (we recall that the light scotogenic dark matter $m_{\rm DM} \lesssim $ MeV may conflict the observed upper limit of ${\rm BR}(\mu \rightarrow e\gamma)$). Although the heavy scotogenic dark matter $m_{\rm DM} \gtrsim 10^9$ GeV is consistent with $\mu \rightarrow  e \gamma$ constraint, the predicted branching ration is too small too observation. 

\begin{figure}[t]
\begin{center}
\includegraphics{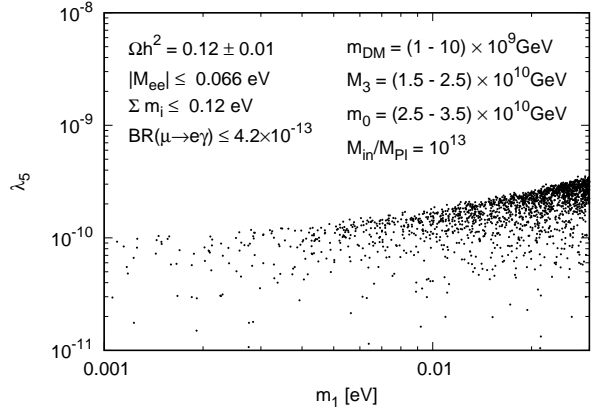}
\caption{The allowed parameter space $(m_1,\lambda_5)$ for scotogenic model with $m_{\rm DM} \sim 10^{9}$ GeV if $M_{\rm in}/M_{\rm Pl} = 10^{13}$ and $\beta > \beta_{\rm c}$.}
\label{fig:Scoto_PBH_allowed_m1_lamblda5_PBH_DOM}
\end{center}
\end{figure}

Figure. \ref{fig:Scoto_PBH_allowed_m1_lamblda5_PBH_DOM} shows that the allowed parameter space $(m_1,\lambda_5)$ for scotogenic model with $m_{\rm DM} \sim 10^{9}$ GeV if $M_{\rm in}/M_{\rm Pl} = 10^{13}$ and $\beta > \beta_{\rm c}$. 

We conclude that more heavy scotogenic dark matter ($m_{\rm DM} \gtrsim$ PeV) is allowed if $m_{\rm DM} \gtrsim 10^9$ GeV for $M_{\rm in}/M_{\rm Pl} \sim 10^{13}$ and there is an additional contribution from the PBH decay; however, a lepton flavor violating process $\mu \rightarrow  e  \gamma$ should be almost completely suppressed. This conclusion is the fourth advantage in this paper.

\section{Summary\label{sec:summary}}
In this paper, we have studied the effect of the scotogenic dark matter on the PBHs and vice versa. In the scotogenic model, the lightest heavy Majorana particle is a good candidate of the dark matter. This scotogenic dark matter produced by the freeze-out mechanism. If the PBHs evaporate after the freeze-out of the dark matter, $t_{\rm FO} < t_{\rm evap}$, then the dark matter particles produced by Hawking radiation of the PBHs may contribute to the final relic abundance of the dark matter. 

In the case of $\beta < \beta_{\rm c}$, the PBHs evaporate in the radiation dominated era, from the requirement of the $t_{\rm FO} < t_{\rm evap}$ and a constraint of the mass of dark matter particles (WDM constraint), we have obtained more stringent upper limit of the initial mass of the PBHs $10^4 \lesssim M_{\rm in}/M_{\rm Pl} \lesssim  10^{10}$. Moreover, due to the existence of the dark matter from freeze-out, the relic abundance of the dark matter produced from PBHs $\Omega_{\rm PBH}  h^2$ should be decrease with $\Omega_{\rm FO}  h^2$. Thus, the upper limit of the initial PBHs density $\beta$ should decrease for coexistence the scotogenic dark matter from freeze-out and the scotogenic dark matter from PBHs. For $\mathcal{O}$ (1) TeV scotogenic dark matter, or equivalently for $(M_{\rm in}/M_{\rm Pl})_{\rm max} \sim 10^{10}$, $\beta \lesssim 5 \times 10^{-11} (M_{\rm in}/M_{\rm Pl})^{-1/2}$ have been obtained for the observed relic abundance of the dark matter, the data from neutrino oscillation experiments and the upper limit of the branching ratio of the $\mu \rightarrow e \gamma$ process.

On the other hand, in the case of $\beta > \beta_{\rm c}$, the PBHs evaporate in the PBH dominated era, the entropy produced by the PBH decay may allow more freedom for the dark matter parameter space in the scotogenic model. We have studied the  possibility of acceptance of more light ($m_{\rm DM} \lesssim$ MeV) or more heavy ($m_{\rm DM} \gtrsim$ PeV) scotogenic dark matter with PBHs. We have shown that the light scotogenic dark matter ($m_{\rm DM} \lesssim$ MeV) conflict the observed upper limit of the branching ration of $\mu \rightarrow e  \gamma$. Thus, we have concluded that more light scotogenic dark matter ($m_{\rm DM} \lesssim$ MeV) should be excluded even if there is an additional contribution from the PBH decay. On the other hand, more heavy scotogenic dark matter ($m_{\rm DM} \gtrsim$ PeV) is allowed if $m_{\rm DM} \gtrsim 10^9$ GeV for $M_{\rm in}/M_{\rm Pl} \sim 10^{13}$ and there is an additional contribution from the PBH decay; however, a lepton flavor violating process $\mu \rightarrow  e \gamma$ should be almost completely suppressed.

\vspace{3mm}







\end{document}